%% file: btoll.tex
\documentclass[12pt]{revtex4}

\usepackage[dvips]{graphics,color}
\begin{document}


\def\lqcd{\Lambda_{\rm QCD}}
\def\Oalphas{${\cal O}(\alpha_s)$}
\def\Oalphas2b0{${\cal O}(\alpha_s^2 \beta_0)$}
\def\xslash#1{{\rlap{$#1$}/}}
\def\dsl{\,\raise.15ex\hbox{/}\mkern-13.5mu D}

\title{Phenomenology of $B \to \pi \pi, \pi K$ Decays at ${\cal O}(\alpha_s^2 \beta_0)$
	in QCD Factorization}

\def\addtoronto{
Department of Physics, University of Toronto\\
60 St.~George Street, Toronto, Ontario,
Canada M5S 1A7\vspace*{6pt}}
\def\addcmu{
Department of Physics, Carnegie Mellon University,\\
Pittsburgh, PA 15213\vspace*{6pt}}

\author{Craig N. Burrell}\affiliation{\addtoronto}
 \email{craig.burrell@utoronto.ca}
\author{Alexander R. Williamson}\affiliation{\addcmu}
 \email{alexwill@andrew.cmu.edu}

\begin{abstract}
We study  ${\cal O}(\alpha_s^2 \beta_0)$ perturbative corrections 
to matrix elements entering two-body exclusive decays of the 
form $\bar{B} \to \pi \pi, \pi K$ in the QCD factorization 
formalism, including chirally enhanced power corrections,
and discuss the effect of these corrections on direct CP asymmetries, which
receive their first contribution at ${\cal O}(\alpha_s)$.  
We find that the ${\cal O}(\alpha_s^2 \beta_0)$ 
corrections are often as large as the ${\cal O}(\alpha_s)$ corrections.
We find large uncertainties due to renormalization scale dependence
as well as poor knowledge of the non-perturbative parameters.
We assess the effect of the perturbative corrections on
the direct CP violation parameters of
$B^0 \to \pi^+ \pi^-$.
\end{abstract}

\maketitle

\pagebreak

\section{Introduction}
In recent years a wealth of new data on two-body nonleptonic $B$ decays 
to light pseudoscalar final states has been produced by
the CLEO \cite{Cronin-Hennessy:kg,Asner:2001eh,Chen:2000hv}, 
BABAR \cite{Aubert:2001hs,Aubert:2001hk,Aubert:2001ap,Aubert:2002ng,Aubert:2002jb,
Aubert:2002jj,Aubert:2002jm,Aubert:2004xc,Aubert:2005av} and 
BELLE \cite{Abe:2001nq,Abe:2001hs,Abe:2004us,Casey:2002yd,Abe:2005dz} experiments.
This experimental program provides a rich context for the precision study of the
weak sector of the standard model.  However, in non-leptonic decays all of the 
final state particles are QCD bound states which interact strongly
with one another.
There is therefore nonperturbative physics in the low energy matrix
elements which is an obstacle to precision calculations.

These low energy matrix elements can be evaluated if it is assumed that they
factorize into simpler matrix elements \cite{Bauer:1986bm}.  For example,
\begin{equation}
\label{naive_factorization}
\langle \pi^+ K^- | (\bar{u}b)_{V-A}(\bar{s}u)_{V-A}|\bar{B}\rangle \to
 \langle K^- |(\bar{s}u)_{V-A}|0 \rangle \langle \pi^+|(\bar{u}b)_{V-A}|\bar{B}\rangle.
\end{equation}
The matrix elements on the right--hand side can be parametrized in 
terms of form factors and decay constants. This
`naive factorization' prescription has in some
cases proven to be a remarkably successful approximation 
\cite{Deandrea:1993ma, NRSX, Neubert:1997uc}.
As it stands, however, there is no way to improve the
calculation by making systematic corrections in a controlled
expansion. 
Moreover, the missing `non-factorizable' physics is
responsible for final-state rescattering and strong interaction phase shifts,
and is therefore of considerable interest.

Beneke, Buchalla, Neubert, and
Sachrajda (BBNS)  \cite{bbns1,bbns2,bbns3}
have argued that for certain classes of two-body nonleptonic $B$ decays
the strong interactions which break factorization are perturbative in the heavy quark
limit \cite{bbns1, bbns2}.
The physical picture behind this claim
is `color transparency': 
gluons must be energetic to resolve the small color dipole structure of the 
energetic light meson in the final state.  The BBNS
proposal, called QCD factorization, was accompanied by a demonstration that,
for heavy-light final states, it holds up to two-loop order \cite{bbns2} in the
heavy quark limit.  This conclusion was subsequently extended
to all orders in perturbation theory \cite{Bauer:2001yt,Bauer:2001cu}.  

The BBNS proposal reproduces naive factorization as the
leading term in an expansion in $\alpha_s$ and $\lqcd/m_b$, thereby placing
naive factorization on a more secure theoretical foundation.
Although there has been some recent progress
\cite{Bauer:2002aj, Chay:2002vy, Chay:2002mw, Beneke:2002ph, Rothstein:2003wh,
Chay:2003ju, Bauer:2004tj},
at present
there exists no systematic way to address the $\lqcd/m_b$ corrections.
The perturbative corrections, on the other hand, can be calculated, and
the ${\cal O}(\alpha_s)$ corrections are known for a variety of decay
modes \cite{bbns1, bbns2, bbns3, Muta:2000ti, Du:2000ff, Du:2001hr}.

In this paper we study ${\cal O}(\alpha_s^2 \beta_0)$ perturbative 
corrections to $B$ decays of the form $B \to \pi \pi,\pi K$.  
Though this is
only a subset of the full ${\cal O}(\alpha_s^2)$ correction, 
the method is motivated by the empirical observation
that the  ${\cal O}(\alpha_s^2\beta_0)$ contribution often dominates the full result.  
For example, 
this is true for $R(e^+ e^- \to \mbox{hadrons} )$ \cite{Rhadrons}, 
$\Gamma(\tau \to \nu_\tau + \mbox{hadrons} )$ \cite{taudecay},
and $\Gamma( b \to X_u e \bar{\nu}_e )$\cite{btoupert}.
The dominance of the ${\cal O}(\alpha_s^2\beta_0)$ contribution 
becomes rigorous in the very formal `large-$\beta_0$ limit' of QCD,
where the number of colors is fixed
and the number of flavours $n_f \to -\infty$, resulting in 
$\beta_0 = 11 - 2/3 n_f \to \infty$.  

The perturbative corrections we consider arise from three sources:
`non-factorizable' vertex corrections, QCD penguin diagrams, and
spectator quark interactions.  In the BBNS framework the computation of 
these amplitudes requires
the introduction of a number of nonperturbative parameters.  
We study the numerical significance of the uncertainties due to these parameters.
We neglect power corrections of the form ${\cal O}(\lqcd/m_b)^n$, with the
exception of a class of `chirally enhanced' corrections that can be 
numerically significant.  Renormalon studies of these decays
indicate that the leading power corrections
from soft gluons are at  ${\cal O}(\lqcd/m_b)$ \cite{Neubert:2002ix}.  
This is in contrast to
$B \to D \pi$ decays where a similar analysis points to leading power
corrections at ${\cal O}(\lqcd/m_b)^2$ \cite{Burrell:2001pf, Becher:2001hu}.

The calculations presented in this paper are similar to calculations performed
earlier by Neubert and Pecjak in \cite{Neubert:2002ix}, although
they differ in several important ways.  The goal of the previous paper was to study
power corrections to $B\to LL$ decays in
a manner similar to what had been done for $B\to D^{(*)}L$ decays in
\cite{Burrell:2001pf, Becher:2001hu}.
The authors calculated the amplitudes to
$\mathcal{O}(1/\beta_0)$, which is subleading in the large-$\beta_0$ limit.
With these expressions they derived
predictions for the CP asymmetries of several decay modes.
The calculation involved summing a class of graphs
to all orders in perturbation theory, and extracting from
their large order perturbative behaviour
information about power corrections.
Estimates of the $\mathcal{O}(\alpha_s^n \beta_0^{n-1})$ corrections were also
made in the large-$\beta_0$ limit.

Our focus is the convergence behaviour of perturbation theory in the
BBNS framework.
Instead of calculating to subleading order in the large-$\beta_0$ limit,
we restrict ourselves to $\mathcal{O}(\alpha_s^2 \beta_0)$ corrections.
To avoid the need for Wilson coefficients evaluated at NNLO, we
concentrate on observables which vanish
at leading order in perturbation theory.
In particular, we study the direct CP
asymmetries for six pseudo-scalar final states ${\cal A}_\mathrm{CP}^\mathrm{Dir}(\pi \pi, \pi K)$.
Though several of these modes also exhibit indirect CP violation, we 
restrict our discussion to direct CP violation only.
We find that the ${\cal O}(\alpha_s^2 \beta_0)$
corrections are similar in size to the ${\cal O}(\alpha_s)$ values.
We also find that the greatest uncertainty in the CP asymmetries
is due to the renormalization scale dependence, which is enhanced
by the ${\cal O}(\alpha_s^2 \beta_0)$ corrections.
In contrast,  the uncertainties induced in the asymmetries
by nonperturbative parameters are relatively small. Of
particular interest is the mode ${\cal A}_\mathrm{CP}^\mathrm{Dir}(\pi^- \bar{K}^0)$ which,
to the order we work, is independent of most of the nonperturbative
parameters in the analysis.

At the end of the paper we present a more detailed analysis 
of the direct asymmetry parameter ${\cal A}_{\pi\pi}$,
which has attracted considerable interest recently \cite{Aubert:2002jb,Abe:2004us}.
We find that this parameter
receives a substantial correction at ${\cal O}(\alpha_s^2 \beta_0)$, 
the size of which we give as a function of the unitarity angle
$\gamma$.   We examine the relationship between this
quantity and the current experimental values, and find them to be in
agreement within the large theoretical and experimental uncertainties.

The structure of this paper is as follows: 
in the first Section we briefly review the theoretical context for
our calculation; in the subsequent Section we give an outline of our method and
collect most of our analytical results.  This is followed by a brief phenomenological
study of direct CP asymmetries for a variety of different decays 
to $\pi \pi, \pi K$ final states, paying special attention to the  
${\cal O}(\alpha_s^2 \beta_0)$ corrections.  In the concluding Section we
summarize our results.

\section{Theoretical Background}
\label{theoretical_background}
We work in the
weak effective theory where the weak bosons and top quark have been integrated out.  
The effective Hamiltonian, valid below $M_W$, is
\begin{equation}
\label{eff_Hamiltonian1}
{\cal H}_{\mathrm{eff}} = \frac{G_F}{\sqrt{2}} \sum_{p=u,c} \lambda_p^{(\prime)} \left( C_1 {\cal O}_1^p + 
	C_2 {\cal O}_2^p + \sum_{i=3,...,6} C_i {\cal O}_i + C_{8g} {\cal O}_{8g} \right) + h.c.
\end{equation}
where the CKM matrix elements are $\lambda_p = V_{ps}^* V_{pb}$ for the
$\Delta S=1$ Hamiltonian and $\lambda_p^\prime = V_{pd}^* V_{pb}$ for decays
to non--strange final states.
The effective operators mediating the decays are divided into left-handed
current-current operators $({\cal O}_{1,2}^p)$, QCD penguin operators
$({\cal O}_{3,...,6})$,
and a chromomagnetic dipole operator $( {\cal O}_{8g})$.  Explicitly, the
operator basis for the $\Delta S = 1$ Hamiltonian is \cite{Buchalla:1995vs}
\begin{eqnarray}
\label{operator_set}
 {\cal O}_1^p &=& (\bar p b)_{V-A} (\bar s p)_{V-A} \,,
    \hspace{2.5cm}
    {\cal O}^p_2 = (\bar p_i b_j)_{V-A} (\bar s_j p_i)_{V-A} \,, \nonumber\\
   {\cal O}_3 &=& (\bar s b)_{V-A} \sum{}_{\!q}\,(\bar q q)_{V-A} \,,
    \hspace{1.7cm}
    {\cal O}_4 = (\bar s_i b_j)_{V-A} \sum{}_{\!q}\,(\bar q_j q_i)_{V-A} \,,
    \nonumber\\
   {\cal O}_5 &=& (\bar s b)_{V-A} \sum{}_{\!q}\,(\bar q q)_{V+A} \,, 
    \hspace{1.7cm}
    {\cal O}_6 = (\bar s_i b_j)_{V-A} \sum{}_{\!q}\,(\bar q_j q_i)_{V+A} \,,
    \nonumber\\
   {\cal O}_{8g} &=& \frac{-g_s}{8\pi^2}\,m_b\,
    \bar s\sigma_{\mu\nu}(1+\gamma_5) G^{\mu\nu} b \, .
\end{eqnarray}
The $\Delta S = 0$ operator set is as above with the $s$ fields replaced by $d$ fields.
In these expressions we use the shorthand $(\bar{q} q')_{V \pm A} = \bar{q} \gamma^\mu (1 \pm \gamma_5) q'$
for the Dirac structures.   Roman indices on quark fields denote $SU(3)$ color structure, 
and the summations over $q$
in the penguin operators ${\cal O}_{3-6}$ run over all five active quark flavours 
$q \in \{d, u, s, c, b \}$.
Some authors include in (\ref{operator_set}) a set of electroweak penguin operators which,
however, produce only small effects and are neglected in our analysis.
The values of the Wilson coefficients $C_i(\mu)$ are obtained by matching the effective theory onto
the full theory at $\mu = m_W$ and running down to $\mu \sim m_b$.
This procedure has been carried out to NLO in QCD, the results of which can be
found in \cite{Buchalla:1995vs}.

 The low energy dynamics are contained in the matrix
elements of the four-quark operators ${\cal O}_i$.
In the BBNS framework these matrix elements are given by
\begin{eqnarray}
\label{master_equation}
\langle M_1 M_2 | {\cal O} | \bar{B} \rangle &=& F^{B \to M_1}(m_{M_2}^2) f_{M_2} \int_0^1 \, dx \, T^I(x) \Phi_{M_2}(x) 
	+ (M_1 \leftrightarrow M_2) \nonumber \\
&+& \int_0^1\,dx\,dy\,d\xi \, T^{II}(x,y,\xi) \Phi_{M_1}(y) \Phi_{M_2}(x) \Phi_{B}(\xi) 
	+ {\cal O}\left( \frac{\lqcd}{m_b} \right).
\end{eqnarray}
where $M_1$ is the meson which receives the
spectator quark of the $B$ meson and $M_2$ is called
the `emission meson'.  The nonperturbative elements in this
expression are the $B$ decay form factors $F^{B \to M}$,
the final state meson decay constants $f_{M}$, and 
the light--cone momentum distribution amplitudes $\Phi_{M}$,
which give the probability for a valence quark to carry a 
particular fraction of the meson's light-cone momentum.
The quark and antiquark composing the emission meson 
$M_2$ are assigned momentum fractions $x$ and $\bar{x}$,
respectively.  Likewise the quark and antiquark in $M_1$ are
assigned momentum fractions $y$ and $\bar{y}$, respectively.  
The light antiquark in the $B$
meson is assigned momentum fraction $\xi$ of the $B$ meson
momentum. In the heavy quark limit, we can neglect components 
of momentum transverse to the light cone, and consider only the 
Fock state containing the valence quark and antiquark. Thus we have
$\bar{x} = 1-x$, and likewise for $y$ and $\xi$.   

The factorization-breaking
corrections are contained in the hard-scattering kernels $T^I(x)$ and $T^{II}(x,y,\xi)$,
each of which has a perturbative expansion.
At leading order, the hard-scattering kernels take the values \cite{bbns1, bbns2, Manohar:1994kq}
\begin{equation}
T^I(x) = 1 + {\cal O}(\alpha_s); \hspace{2cm} T^{II}(x,y,\xi) = 0 + {\cal O}(\alpha_s)
\end{equation}
and, given that the light--cone wavefunctions $\Phi_{M_i}$ are normalized to unity,
(\ref{master_equation}) reduces to naive factorization.  

 The nonperturbative light--cone distribution amplitudes (LCDAs) $\Phi_M$ in (\ref{master_equation})
are defined by
\cite{Braun:1988qv,Braun:1989iv}
\begin{eqnarray}
\label{LCDAdefs}
\langle P(p) | \bar{q}_\beta (z_2)q_\alpha(z_1) | 0 \rangle &=& \\
	&\,& \hspace{-2.5cm} i \frac{f_P}{4}
	\int_0^1 dx e^{i(x\,p\cdot z_2 + \bar{x}\,p \cdot z_1)} 
	\left[ \xslash{p} \gamma_5 \Phi(x) - \mu_P \gamma_5 \left( \Phi_p(x) - \sigma^{\mu\nu}p_\mu (z_2-z_1)_\nu
	\frac{\Phi_\sigma(x)}{6} \right) \right]_{\alpha\beta} \nonumber .
\end{eqnarray}
In this equation $\Phi(x)$ is the meson twist--2 LCDA, and $\Phi_{p,\sigma}(x)$ are twist--3 LCDAs
which will contribute to the `chirally enhanced' power corrections below.  The quantity $\mu_P$ 
appearing in (\ref{LCDAdefs}) is a `chiral enhancement' factor
\begin{equation}
\label{chiral}
\mu_P = \frac{m_P^2}{m_q + m_{\bar{q}}}
\end{equation}
where $q$ and $\bar{q}$ are the quarks which comprise the valence state of the pseudoscalar meson $P$.
In practice one introduces these LCDAs into Feynman amplitudes by replacing quark bilinears with a 
projection matrix $M$
\cite{bbns3}
\begin{equation}
\bar{u}_{\beta,a}(x p)\Gamma_{\beta \alpha, a b} v_{\alpha,b}(\bar{x}p) \to 
	\frac{i f_P}{4 N_c} \int_0^1 dx M^P_{\alpha \beta}\Gamma_{\beta \alpha, a a}
\end{equation}  
where
\begin{equation}
\label{twistproj}
M^P = \xslash{p}\gamma_5 \Phi(x) - 
	\mu_P \gamma_5 \frac{\xslash{p}_2 \xslash{p}_1}{p_1 \cdot p_2} \Phi_p(x).
\end{equation}
In these expressions Greek indices denote Dirac structure, Roman indices denote
color structure, and $\Gamma$ is an arbitrary combination of Dirac and color matrices.  
The momenta $p_{1,2}$ are the momenta of
the meson quark and antiquark, respectively.  In the collinear limit $p_1 = x p$
and $p_2 = \bar{x} p$.  In order to arrive at (\ref{twistproj})
from the definition (\ref{LCDAdefs}) the equations of motion for the twist--3 LCDAs
and an integration by parts have been 
used to eliminate $\Phi_\sigma$ \cite{bbns3,Beneke:2002bs}.

Throughout this
paper we write the twist--2 LCDAs as a decomposition over Gegenbauer polynomials as
\cite{Lepage:1980fj}
\begin{equation}
\label{wfdecomposition}
\Phi_M(x) = 6 x (1-x) \left[ 1 + \sum_{n=1}^\infty \alpha_n^M(\mu) C_n^{3/2}(2x-1) \right]
\end{equation}       
where the Gegenbauer polynomials $C_n^{3/2}(y)$ are defined by the generating function
\begin{equation}
C_n^{3/2}(y) = \left. \frac{1}{n!} \frac{d^n}{dh^n} (1 - 2 h y + h^2)^{-3/2}\right|_{h=0}.
\end{equation}
The Gegenbauer moments $\alpha_i^M$ have been studied using nonperturbative methods in
QCD and, for many light mesons, estimates exist for the leading moments 
$\alpha_{1,2}^M$ \cite{Braun:1988qv,Chernyak:1983ej,rady}. In the far ultraviolet 
$\mu \to \infty$ we have $\alpha_i^M \to 0$, so at the scale $\mu \sim m_b$, which
is still large compared to the nonperturbative scale of QCD, the Gegenbauer moments
$\alpha_i^M$ are expected to be small.  This statement will be made more
quantitative in Section \ref{phenomenology}.
In the approximation of including only `chirally enhanced' twist--3 contributions, 
the twist--3 LCDA equations of motion constrain $\Phi_p(x)$ to take its asymptotic form
$\Phi_p(x)=1$ \cite{bbns3}.
  
Following the authors of \cite{bbns3} we use the factorization formula (\ref{master_equation})
to rewrite matrix elements of (\ref{eff_Hamiltonian1}) in the convenient form
\begin{equation}\label{Top}
   \langle \pi K | {\cal H}_{\rm eff}| \bar B \rangle
   = \frac{G_F}{\sqrt2} \sum_{p=u,c} \lambda_p\,
   \langle\pi K|{\cal T}_p | \bar B\rangle \,,
\end{equation}
where
\begin{eqnarray}\label{Toper}
   {\cal T}_p &=& a_1\,\delta_{pu}\,
    (\bar u b)_{V-A} \otimes (\bar s u)_{V-A} \nonumber\\
   &+& a_2\,\delta_{pu}\,
    (\bar s b)_{V-A} \otimes (\bar u u)_{V-A} \nonumber\\
   &+& a_3 \sum{}_{\!q}\, (\bar s b)_{V-A} \otimes
    (\bar q q)_{V-A} \nonumber\\
   &+& a_4^p \sum{}_{\!q}\, (\bar q b)_{V-A} \otimes
    (\bar s q)_{V-A} \nonumber\\
   &+& a_5 \sum{}_{\!q}\, (\bar s b)_{V-A} \otimes
    (\bar q q)_{V+A} \nonumber\\
   &+& a_6^p \sum{}_{\!q}\, (-2)(\bar q b)_{S-P} \otimes
    (\bar s q)_{S+P} 
\end{eqnarray}
where $(\bar{q} q')_{S\pm P}=\bar{q} (1\pm\gamma_5) q'$, and a
summation over $q \in \{u,d\}$ is implied.  There is a similar expression
for $\pi\,\pi$ final states, obtained by replacing the $s$ quark by a $d$
quark.  Matrix elements of
operators containing the $\otimes$ product are to be evaluated
as one would in naive factorization
\begin{equation}
\langle M_1 M_2 | j_1 \otimes j_2 | B \rangle \equiv  \langle M_1 | j_1 | B \rangle \, \langle M_2 | j_2| 0 
	\rangle \mbox{ or }  \langle M_2 | j_1 | B \rangle \, \langle M_1 | j_2| 0 \rangle
\end{equation}
where the choice depends on the specific quark content of the mesons in the process under consideration.
The nonfactorizable corrections are contained in the coefficients 
$a_i$. Though we have not explicitly indicated it in (\ref{Toper}), in general these
coefficients are mode specific, dependent on the shapes of the LCDAs of the final
state particles.  We present the explicit forms for $a_i$ in section
\ref{perturbative_corrections}.

\section{Perturbative Corrections}
\label{perturbative_corrections}
We consider three classes of diagrams:  factorization-breaking vertex diagrams,
strong interactions with the initial state spectator quark, and QCD penguin diagrams.
They are shown in Figures 
\ref{fig:vertex_diagrams} -- \ref{fig:penguin_diagrams}.
The first and third
of these classes contribute to the hard scattering kernel $T^I$ in
(\ref{master_equation}); the second contributes to $T^{II}$.
In this paper we do not include the power suppressed weak annihilation 
diagrams which have been studied by other authors \cite{bbns3}.

\begin{figure}[htbp]
\centerline{\scalebox{0.6}{\includegraphics{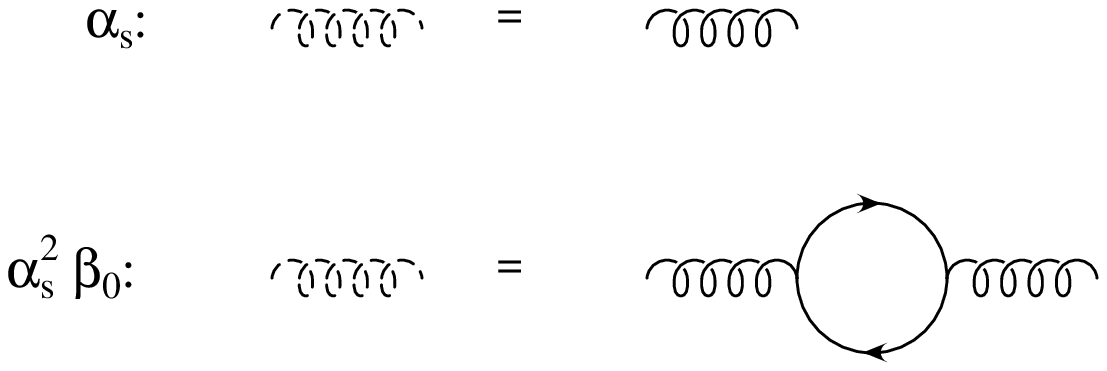}}}
\caption{The form of the dashed gluon line in Figures \ref{fig:vertex_diagrams} -- \ref{fig:penguin_diagrams}
at each order in perturbation theory.  Where an undressed gluon line provides the ${\cal O}(\alpha_s)$
contribution, the fermion--loop self--energy correction produces the 
${\cal O}(\alpha_s^2 \beta_0)$ contributions.}
\label{fig:gluons}
\end{figure}

The diagrams at ${\cal O}(\alpha_s^2 \beta_0)$  are obtained by replacing 
the gluon in the ${\cal O}(\alpha_s)$ diagram by a gluon with a fermion--loop self--energy
correction, as shown in Figure \ref{fig:gluons}, followed by the replacement $n_f \to -3\beta_0/2$. 

All of the perturbative corrections
are contained in the coefficients $a_i$ defined in (\ref{Toper}).
These coefficients may be written as
\begin{eqnarray}\label{ai}
   a_{1} &=& C_1 + \frac{C_2}{N_c} \Bigg[ 1 + C_F \left\{ \frac{\alpha_s(\mu)}{4\pi}\,V_{M_2}
	+ \left( \frac{\alpha_s(\mu)}{4\pi} \right)^2 \beta_0 \tilde{V}_{M_2} \right\} \nonumber \\
	&\,& \hspace{3cm} + \frac{4 C_F \pi^2}{N_c} \left\{ \frac{\alpha_s(\mu_h)}{4\pi}\,
    H_{M_2 M_1} +  \left( \frac{\alpha_s(\mu_h)}{4\pi} \right)^2 \beta_0 \tilde{H}_{M_2 M_1} \right\} \Bigg],
    \nonumber\\
   a_{2} &=& C_2 + \frac{C_1}{N_c} \Bigg[1+ C_F \left\{ \frac{\alpha_s(\mu)}{4\pi}\,V_{M_2}
	+ \left( \frac{\alpha_s(\mu)}{4\pi} \right)^2 \beta_0 \tilde{V}_{M_2} \right\} \nonumber \\
	&\,& \hspace{3cm} + \frac{4 C_F \pi^2}{N_c} \left\{ \frac{\alpha_s(\mu_h)}{4\pi}\,
    H_{M_2 M_1} +  \left( \frac{\alpha_s(\mu_h)}{4\pi} \right)^2 \beta_0 \tilde{H}_{M_2 M_1} \right\} \Bigg],
    \nonumber\\
   a_{3} &=& C_3 + \frac{C_4}{N_c} \Bigg[ 1 + C_F \left\{ \frac{\alpha_s(\mu)}{4\pi}\,V_{M_2}
	+ \left( \frac{\alpha_s(\mu)}{4\pi} \right)^2 \beta_0 \tilde{V}_{M_2} \right\} \nonumber \\
	&\,& \hspace{3cm} + \frac{4 C_F \pi^2}{N_c} \left\{ \frac{\alpha_s(\mu_h)}{4\pi}\,
    H_{M_2 M_1} +  \left( \frac{\alpha_s(\mu_h)}{4\pi} \right)^2 \beta_0 \tilde{H}_{M_2 M_1} \right\} \Bigg],
    \nonumber\\
   a_{4}^p &=& C_4 + \frac{C_3}{N_c} \Bigg[ 1 + C_F \left\{ \frac{\alpha_s(\mu)}{4\pi}\,V_{M_2}
	+ \left( \frac{\alpha_s(\mu)}{4\pi} \right)^2 \beta_0 \tilde{V}_{M_2} \right\}
\nonumber \\
&\,& \hspace{3cm} + \frac{4 C_F \pi^2}{N_c} \left\{ \frac{\alpha_s(\mu_h)}{4\pi}\, H_{M_2 M_1}
	+  \left( \frac{\alpha_s(\mu_h)}{4\pi} \right)^2 \beta_0 \tilde{H}_{M_2 M_1} \right\} \Bigg]
\nonumber \\ 
&\,& \hspace{3cm}
 + \frac{C_F}{N_c} \Bigg\{ \frac{\alpha_s(\mu)}{4\pi}  P_{M_2,2}^p 
	+ \left( \frac{\alpha_s(\mu)}{4\pi} \right)^2 \beta_0 \tilde{P}_{M_2,2}^p \Bigg\},
    \nonumber \\
   a_{5} &=& C_5 + \frac{C_6}{N_c} \Bigg[ 1 - C_F \left\{ \frac{\alpha_s(\mu)}{4\pi}\, V'_{M_2}   +
	\left( \frac{\alpha_s(\mu)}{4\pi} \right)^2 \beta_0 \tilde{V}'_{M_2} \right\} \nonumber \\
	&\,& \hspace{3cm} - \frac{4 C_F \pi^2}{N_c} \left\{ \frac{\alpha_s(\mu_h)}{4\pi}\,H'_{M_2 M_1} +
	\left( \frac{\alpha_s(\mu_h)}{4\pi} \right)^2 \beta_0 \tilde{H}'_{M_2 M_1} \right\} \Bigg] ,
    \nonumber \\
   a_{6}^p &=& C_6 + \frac{C_5}{N_c} \Bigg[ 1 + C_F \left\{ \frac{\alpha_s(\mu)}{4\pi} \, (-6)
	+ \left( \frac{\alpha_s(\mu)}{4\pi} \right)^2 \beta_0 (-4) \right\} \Bigg] \nonumber \\
    &\,& \hspace{3cm} + \frac{C_F}{N_c} \left\{ \frac{\alpha_s(\mu)}{4\pi}  P_{M_2,3}^p + \left( \frac{\alpha_s(\mu)}{4\pi} \right)^2 \beta_0 \tilde{P}_{M_2,3}^p \right\}\,,
\end{eqnarray}
where $C_i\equiv C_i(\mu)$.
The nonperturbative physics is contained in the functions labelled
$V,H,$ and $P$, according to the type of diagram from which
they arise.   In particular, the non-factorizable vertex corrections,
treated in section \ref{sec:vertex_diagrams} below, produce the functions $V_M^{(\prime)}$
and  $\tilde{V}_M^{(\prime)}$.  The scattering of hard gluons off
the spectator quark, treated in section \ref{sec:hs_diagrams}, gives rise to
the functions $H_{M_2 M_1}^{(\prime)}$ and $\tilde{H}_{M_2 M_1}^{(\prime)}$.
Graphs with penguin topologies,
discussed in section \ref{sec:penguin_diagrams},
produce the $P_{M_2,n}^p$ and
$\tilde{P}_{M_2,n}^p$ functions, where $n$ refers to the twist of the LCDA
that enters the evaluation of the function.
All of these functions consist of convolutions of hard-scattering kernels with meson
light--cone distribution amplitudes, as will be shown below.

The scale at which renormalization scale dependent quantities
are to be evaluated differs depending on the source of the contribution.
In particular,  while the vertex and penguin diagrams are evaluated
at a scale $\mu \sim m_b$, the spectator scattering contributions
are evaluated at a lower scale $\mu_h \sim \sqrt{m_b \lqcd}$.  This applies to all
scale dependent quantities multiplying the spectator scattering functions
$H^{(\prime)}$, including the Wilson coefficients \cite{bbns3}.

\subsection{ Vertex Diagrams }
\label{sec:vertex_diagrams}
\begin{figure}[htbp]
\centerline{\scalebox{0.6}{\includegraphics{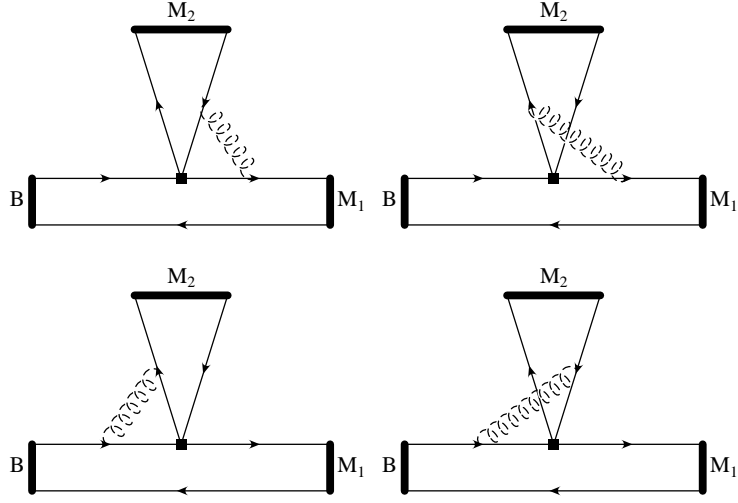}}}
\caption{The factorization-breaking vertex diagrams.}
\label{fig:vertex_diagrams}
\end{figure}

The first class of diagrams we consider are those shown in Figure \ref{fig:vertex_diagrams} in which
a hard gluon is exchanged between the emission meson $M_2$ and the quarks involved
in the $B \to M_1$ transition.   These amplitudes are proportional to both $f_{M_2}$
and $F_0^{B \to M_1}(0)$ and contribute to the kernel $T^I$ in (\ref{master_equation}).
In terms of the coefficients $a_i$ in (\ref{ai}) they produce $V^{(')}_{M_2}$ at
${\cal O}(\alpha_s)$ and $\tilde{V}^{(')}_{M_2}$ at ${\cal O}(\alpha_s^2 \beta_0)$.

 Although when evaluated in $D=4-2\,\epsilon$ dimensions
each of these diagrams contains a $1/\epsilon^2$ pole from
infrared and collinear divergences,
these cancel in the sum of the four diagrams, leaving a residual UV $1/\epsilon$ pole to be
renormalized.  We renormalize in the $\overline{\mbox{MS}}$ scheme, treating $\gamma_5$ in the naive
dimensional regularization (NDR) prescription \cite{Buras:1989xd}.
For completeness we restate the result of Ref. \cite{bbns3} for the ${\cal O}(\alpha_s)$ contributions
\begin{equation}
V_M =  -6 \left[ \ln \left(\frac{\mu^2}{m_b^2}\right)+\frac{5}{3} \right] - 1 +
	\int_0^1 \; dx \; g(x) \Phi_M(x)
\end{equation}
and
\begin{equation}
V'_M = -6 \left[ \ln \left(\frac{\mu^2}{m_b^2}\right) + \frac{5}{3} \right] + 11 + \int_0^1 \; dx \; g(1-x) \Phi_M(x)
\end{equation}
where the integration kernel is
\begin{eqnarray}
g(x) &=& \left[ \frac{3 (1-2x)}{2 (1-x)} \ln(x) - \frac{1}{2}\left( 7 + 3 i \pi \right) + 
	( x \leftrightarrow \bar{x} ) \right]  \nonumber \\
   &+& \left[ \frac{\ln(x) }{2 (1-x)} - 2 i \pi \ln(x) - \ln^2(x)- 2 \mbox{Li}_2\left(1 - x\right) -
	( x \leftrightarrow \bar{x} ) \right].
\end{eqnarray}

For the next order result we find
\begin{eqnarray}
\tilde{V}_M &=& -3 \left[ \ln\left(\frac{\mu^2}{m_b^2}\right) + \frac{5}{3}  \right]^2 + 
	\left[ \ln\left(\frac{\mu^2}{m_b^2}\right) + \frac{5}{3}\right] \int_0^1 \; dx \; g(x) \Phi_M(x) \nonumber \\
	&+& \int_0^1 \; dx \; h(x) \Phi_M(x) - \frac{65}{12}
\end{eqnarray}
and
\begin{eqnarray}
\tilde{V}'_M &=& -3 \left[ \ln\left(\frac{\mu^2}{m_b^2}\right) + \frac{5}{3}  \right]^2 + 
	\left[ \ln\left(\frac{\mu^2}{m_b^2}\right)+\frac{5}{3}  \right] \int_0^1 \; dx \; g(1-x) \Phi_M(x) \nonumber \\
	&+& \int_0^1 \; dx \; h(1-x) \Phi_M(x) - \frac{5}{12}.
\end{eqnarray}

 The function $h(x)$ appearing in this expression is given by
\begin{eqnarray}
h(x) &=& \left[ -\frac{3 (1 - 3 x)}{4 (1-x)} \ln^2(x) + \left( \frac{7 (1- 2 x)}{4 (1-x)}  + \frac{3}{2} i \pi \right) \ln(x) +
	\frac{ 3 x \mbox{Li}_2(1-x)}{2 (1-x)} \right.\nonumber \\
&\;& \hspace{2cm} -\frac{1}{4} \left( 15 + 7 i \pi \right) + ( x \leftrightarrow \bar{x} ) \Bigg] \nonumber \\
&+&
	\left[ \ln^3(x) - \left( \frac{5-3 x}{4 (1-x)} + 2 \ln(1-x) - i \pi \right) \ln^2(x) + 
	\left( \frac{1}{4 (1-x)} + 2 \pi^2 - \frac{3}{2} i \pi \right) \ln(x) \right. \nonumber \\
&\;& \hspace{2cm} -\left. \frac{4 - 3 x}{2 (1-x)} \mbox{Li}_2(1-x) -
	2 \mbox{Li}_3(1-x) - 4 \mbox{Li}_3(x) -  ( x \leftrightarrow \bar{x} ) \right].
\end{eqnarray}
This function has previously been derived in \cite{Becher:2001hu}; we confirm that result.

One may notice from a comparison of the factorized operator (\ref{Toper}) with the pattern of vertex graph 
contributions to the $a_i$ coefficients in (\ref{ai}) that the unprimed functions $V,\tilde{V}$ 
are associated with $(V-A) \otimes (V-A)$ operator structures, while $V^\prime,\tilde{V}^\prime$ are
associated with $(V-A) \otimes (V+A)$ structures.  The remaining operator structure present in (\ref{Toper})
is $(S-P) \otimes (S+P)$, and this receives a nonzero vertex contribution only when the twist--3 
LCDAs $\Phi_p$ are included. We have used the fact that, in the approximation of including only
the `chirally enhanced' terms at twist--3, $\Phi_p$ has its asymptotic form $\Phi_p(x)=1$
to carry out the momentum fraction integrals, resulting in the constants `$-6$' and `$-4$' appearing
in $a_6$ of (\ref{ai}) at ${\cal O}(\alpha_s)$ and ${\cal O}(\alpha_s^2 \beta_0)$, respectively.   

Using the Gegenbauer expansion for the LCDAs $\Phi_M$, we carry out the 
integration over momentum fraction $x$ to obtain
\begin{eqnarray}
\int_0^1 dx \; g(x) \Phi_M(x) &=& - \frac{15}{2} - 3 i \pi
	+ \left(\frac{11}{2} - 3 i \pi \right) \alpha_1^M \nonumber \\
&\;& \hspace{3cm}- \frac{21}{20} \alpha_2^M +
	\left( \frac{79}{36} - \frac{2}{3} i \pi \right) \alpha_3^M + \cdot\cdot\cdot
\end{eqnarray}
and
\begin{eqnarray}
\int_0^1 \; dx \; h(x) \Phi_M(x) &=& \pi^2 - \frac{33}{2} - 6 i \pi
	+ \left( 3 \pi^2 + \frac{65}{6} - \frac{11}{2} i \pi  \right) \alpha_1^M \nonumber \\
	&\;& \hspace{-3cm} + \left( \frac{3}{2} \pi^2 - \frac{7359}{400} - \frac{9}{10} i \pi \right) \alpha_2^M
	+ \left( \frac{5}{18} \pi^2 + \frac{10481}{720} - \frac{37}{15} i \pi \right) \alpha_3^M + \cdot\cdot\cdot
\end{eqnarray}

Thus the vertex diagrams introduce complex phases into the 
amplitude, and the magnitude of the phase depends on the shape of the LCDAs 
parametrized by $\alpha_i^M$.

\subsection{ Spectator Scattering Diagrams }
\label{sec:hs_diagrams}

\begin{figure}[htbp]
\centerline{\scalebox{0.6}{\includegraphics{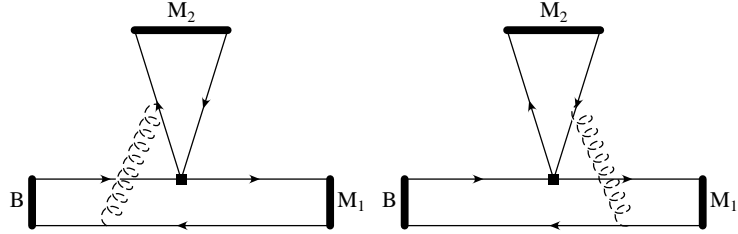}}}
\caption{The factorization-breaking spectator scattering diagrams.}
\label{hs_diagrams}
\end{figure}

The two diagrams involving hard scattering with the spectator quark are shown in Figure 
\ref{hs_diagrams}.
The ${\cal O}(\alpha_s)$ corrections are expressed in Ref. \cite{bbns3} in terms of two 
functions $H^{(')}_{M_2 M_1}$ as shown in (\ref{ai}).  We choose to write these 
functions as
\begin{eqnarray}
\label{Hxyz}
H^{(')}_{M_2 M_1} &=& \frac{ f_B f_{M_1} }{m_B^2 F_0^{B \to M_1}(0) }  \int_0^1 dx \int_0^1 dy \int_0^1 d\xi \left[ 
	_2 H^{(')}_{M_2 M_1}(x,y,\xi) \Phi_{M_2}(x) \Phi_{M_1}(y) \Phi_B(\xi) \right. \nonumber \\
	&\,& \hspace{2cm} + \left. \frac{2 \mu_{M_1}}{m_b}  \,_3 H^{(')}_{M_2 M_1}(x,y,\xi) \Phi_{M_2}(x) \Phi_p^{M_1}(y) \Phi_B(\xi) \right]   .
\end{eqnarray}
where we have divided the integration kernel into twist--2 $( _2 H^{(')} )$ and twist--3 $(   _3  H^{(')} )$ components.
In the approximation $\xi \ll x,y$, which one expects to be valid through most of phase space, the integration kernels are
\begin{eqnarray}
_2 H_{M_2 M_1}(x,y,\xi) &=& _3 H'_{M_2 M_1}(x,y,\xi)  = \frac{1}{\bar{x} \; \bar{y} \; \xi}, \nonumber \\
_2 H'_{M_2 M_1}(x,y,\xi) &=& _3 H_{M_2 M_1}(x,y,\xi)  =\frac{1}{ x \; \bar{y} \; \xi}. 
\end{eqnarray}
If one replaces the light-cone distribution functions $\Phi_M$ with their expansions in terms of Gegenbauer
polynomials (\ref{wfdecomposition}) and carries out the integrations in (\ref{Hxyz}), one finds
\begin{eqnarray}
H^{(')}_{M_2 M_1} &=& \frac{ f_B f_{M_1} }{m_B \lambda_B F_0^{B \to M_1}(0) } \bigg[ 9 ( 1 + \alpha_1^{M_1} + \alpha_2^{M_1} + \cdot \cdot \cdot ) 
	( 1 \pm \alpha_1^{M_2} + \alpha_2^{M_2} \pm \cdot \cdot \cdot )  \nonumber \\
	&+& \frac{6 \mu_{M_1}}{m_b} X_H^{M_1} ( 1 \mp \alpha_1^{M_2} +  \alpha_2^{M_2} \mp \cdot \cdot \cdot ) \bigg]
\end{eqnarray}
where the ellipses denote higher order Gegenbauer moments, and in `$\pm/\mp$' the top symbol applies to $H$ and the bottom symbol to
$H'$.  Following Refs. \cite{bbns1,bbns2,bbns3} we have also introduced two 
parameters $\lambda_B$ and $X_H^{M_1}$ defined by
\begin{equation}
\label{lambda_B}
\int_0^1 d\xi \frac{\Phi_B(\xi)}{\xi} \equiv \frac{m_B}{\lambda_B}; \hspace{2cm} \int_0^1 dy \frac{\Phi_p^{M_1}(y)}{\bar{y}} \equiv X_H^{M_1}.
\end{equation}
Because the light quark in the $B$ meson carries a small momentum fraction, the wavefunction
$\Phi_B(\xi)$ has support only for $0 < \xi \lesssim \lqcd/m_b$.  The definition (\ref{lambda_B}) then
implies $\lambda_B \sim \lqcd$.   
It is necessary to introduce the parameter $X_H^{M_1}$ because, with the asymptotic form for $\Phi_p^{M_1} = 1$, the 
integration contains a logarithmic divergence when the $B$ meson spectator quark enters 
$M_1$ as a soft quark 
$\bar{y} \sim 0$.  This divergence is a consequence of our having neglected 
the small transverse components of 
momentum and quark off-shellness \cite{bbns3}.  $X_H^{M_1}$ is
therefore a new complex nonperturbative parameter, and by 
power counting it is of size $X_H^{M_1} \sim \ln(m_b/\lqcd)$.

At next perturbative order one finds two new functions $\tilde{H}^{(')}_{M_2 M_1}$ defined by
\begin{eqnarray}
\tilde{H}^{(')}_{M_2 M_1} &=& \left[ \ln\left(\frac{\mu^2}{m_B^2}\right) + \frac{5}{3}  \right]  H^{(')}_{M_2 M_1}  \nonumber \\
	&-&  \frac{ f_B f_{M_1} }{m_B^2 F_0^{B \to M_1}(0) }  \int_0^1 dx \int_0^1 dy \int_0^1 d\xi \bigg[ 
	\,_2 \tilde{H}^{(')}_{M_2 M_1}(x,y,\xi) \Phi_{M_2}(x) \Phi_{M_1}(y) \Phi_B(\xi)  \nonumber \\
	&+& \frac{2 \mu_{M_1}}{m_b}  \,_3 \tilde{H}^{(')}_{M_2 M_1}(x,y,\xi) \Phi_{M_2}(x) \Phi_p^{M_1}(y) \Phi_B(\xi) \bigg] 
\end{eqnarray}
where the new hard scattering kernels are
\begin{eqnarray}
_2 \tilde{H}_{M_2 M_1}(x,y,\xi) &=& _3 \tilde{H}'_{M_2 M_1}(x,y,\xi)  = \frac{\ln(\xi \; \bar{y})}{\bar{x} \; \bar{y} \; \xi}, \nonumber \\
_2 \tilde{H}'_{M_2 M_1}(x,y,\xi) &=& _3 \tilde{H}_{M_2 M_1}(x,y,\xi)  =\frac{\ln(\xi \; \bar{y})}{ x \; \bar{y} \; \xi}. 
\end{eqnarray}
Carrying out the integrations explicitly one arrives at
\begin{eqnarray}
\label{tildeH}
\tilde{H}^{(')}_{M_2 M_1} &=& \left[ \ln\left(\frac{\mu^2}{m_B^2}\right) + \frac{5}{3}  \right]  H^{(')}_{M_2 M_1}  \nonumber \\
	&+& \frac{ f_B f_{M_1} }{m_B F_0^{B \to M_1}(0) } \Bigg[ \frac{27}{2 \lambda_B} \bigg( 1 + \frac{17}{9} \alpha_1^{M_1} + \frac{43}{18} \alpha_2^{M_1} + \cdot\cdot\cdot \bigg) 
	\bigg( 1 \pm \alpha_1^{M_2} + \alpha_2^{M_2} \pm \cdot \cdot \cdot \bigg)  \nonumber \\
	&-& \frac{9}{\tilde{\lambda}_B} \left( 1 + \alpha_1^{M_1} + \alpha_2^{M_1} + \cdot \cdot \cdot \right) 
	\left( 1 \pm \alpha_1^{M_2} + \alpha_2^{M_2} \pm \cdot \cdot \cdot \right)  \nonumber \\
	&-& \frac{ 6 \mu_{M_1}}{m_b} \bigg\{ \frac{ \tilde{X}_H^{M_1} }{ \lambda_B} + \frac{X_H^{M_1}}{ \tilde{\lambda}_B} \bigg\}
	\left( 1 \mp \alpha_1^{M_2} + \alpha_2^{M_2} \mp \cdot \cdot \cdot \right)  \Bigg]. 
\end{eqnarray}
The large coefficients in the second line of (\ref{tildeH}) result from the integral $\int_0^1 \ln\bar{y} \, \Phi(y)/\bar{y} \; dy$. 
In the Gegenbauer expansion of the LCDA $\Phi$ all of the Gegenbauer moments $\alpha_i$ enter with large
coefficients, so only if the moments themselves decrease quickly will this integral be well represented by the terms we retain.
In addition we have been forced to introduce two additional parameters similar to those in (\ref{lambda_B}):
\begin{equation}
\label{tildelambda_B}
\int_0^1 d\xi \frac{\ln \xi \; \Phi_B(\xi)}{\xi} \equiv \frac{m_B}{\tilde{\lambda}_B}; \hspace{2cm} 
\int_0^1 dy \frac{\ln \bar{y} \; \Phi_p^{M_1}(y)}{\bar{y}} \equiv \tilde{X}_H^{M_1}.
\end{equation}
By power counting these parameters are of order $\tilde{\lambda}_B \sim \lqcd / \ln(\lqcd/m_b)$ and
$\tilde{X}_H^{M_1} \sim \mbox{Li}_2(-m_b/\lqcd)$.

\subsection{ QCD Penguin Diagrams }
\label{sec:penguin_diagrams}

\begin{figure}[htbp]
\centerline{\scalebox{0.6}{\includegraphics{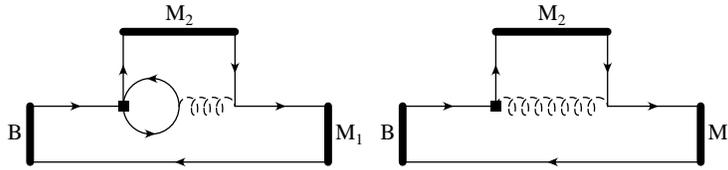}}}
\caption{The penguin and magnetic dipole diagrams.}
\label{fig:penguin_diagrams}
\end{figure}

An important source of strong phases in the decay amplitudes are the QCD penguin diagrams, shown
in Figure \ref{fig:penguin_diagrams}. These diagrams give rise to the functions $P^p_{M,i}$ and
$\tilde{P}^p_{M,i}$ appearing in (\ref{ai}).   The four quark operators in the Hamiltonian 
(\ref{operator_set}) contribute to the left--hand diagram, while the
chromomagnetic dipole operator ${\cal O}_{8g}$ contributes in the right--hand
diagram.  

We begin by stating
the results at ${\cal O}(\alpha_s)$.   At twist--2 one finds \cite{bbns3}
\begin{eqnarray}
\label{PMqcd}
   P_{M,2}^p &=& \int_0^1 dx \, P_{2}^p(x)  \Phi_M(x) \nonumber \\
   P_{2}^p(x) &=& C_1 \left[ \frac43 \ln\frac{m_b}{\mu}
    + \frac23 - G(s_p, x) \right]
    + C_3 \left[ \frac83\ln\frac{m_b}{\mu} + \frac43
    - G(0, x) - G(1, x) \right] \nonumber\\
   &&\mbox{}+ (C_4+C_6) \left[ \frac{4n_f}{3}\ln\frac{m_b}{\mu}
    - (n_f-2) G(0, x) - G(s_c, x) - G(1, x) \right] \nonumber\\
   &&\mbox{}- 2 C_{8g}^{\rm eff} \frac{1}{1-x}
\end{eqnarray}
where $n_f=5$ is the number of active quark flavours, 
$s_q=(m_q/m_b)^2$, and $C_{8g}^{\rm eff} = C_{8g} + C_5$.

The function $G(s, x)$  in (\ref{PMqcd}) is given by the integral
\begin{equation}
G(s,x) = -4\int_0^1\!du\,u(1-u) \ln[s - u (1-u) (1-x) - i\epsilon].
\end{equation}
The integral $\int_0^1\,dx\,G(s_p,x) \Phi(x)$  
is complex and contributes to the strong phase of the amplitude for $s < 1/4$; 
that is, for all quark flavours except the $b$ quark.

The order ${\cal O}(\alpha_s^2 \beta_0)$ results at twist--2 we find to be
\begin{equation}
\label{PtildeMqcd}
\tilde{P}_{M,2}^p = \left[ \ln\left(\frac{\mu^2}{m_b^2}\right) + \frac{5}{3} \right]  P_{M,2}^p 
- \int_0^1 dx \, \ln( x-1-i \epsilon) P_{2}^p(x)  \Phi_M(x).
\end{equation} 

A similar situation exists when one turns to the twist--3 terms.  One finds the leading corrections involve
\begin{eqnarray}
\label{hatPM}
   P_{M,3}^p &=& \int_0^1 dx \, P_{3}^p(x) \Phi^M_p(x)   \\
P_{3}^p(x) &=& C_1 \left[ \frac43\ln\frac{m_b}{\mu}
    + \frac23 - G(s_p, x) \right]
    + C_3 \left[ \frac83\ln\frac{m_b}{\mu} + \frac43
    - G(0, x) - G(1, x) \right] \nonumber\\
   &+& (C_4+C_6) \left[ \frac{4n_f}{3}\ln\frac{m_b}{\mu}
    - (n_f-2) G(0, x) - G(s_c, x) - G(1, x) \right]
    - 2 C_{8g}^{\rm eff} \nonumber
\end{eqnarray}
where the twist--3 distribution function $\Phi^M_p(x)$ has replaced
the twist--2 distribution in (\ref{PMqcd}).

The ${\cal O}(\alpha_s^2 \beta_0)$ function is, in a manner closely 
analogous to (\ref{PtildeMqcd}),
given by
\begin{equation}
\label{hatPtildeMqcd}
\tilde{P}_{M,3}^p = \left[ \ln\left(\frac{\mu^2}{m_b^2}\right) + \frac{5}{3} \right]  P_{M,3}^p 
	- \int_0^1 dx \, \ln( x-1-i \epsilon ) P_{3}^p(x)  \Phi^M_p(x).
\end{equation}

 The expressions which result from carrying out the integrations over momentum
fractions are quite complicated for the penguin diagrams, and we refrain from
presenting them here.

\section{Phenomenological Analysis}
\label{phenomenology}

In this section we take the analytic results of section
\ref{perturbative_corrections} and study direct
CP asymmetries for various $\pi \pi$ and $\pi K$ final states. We 
begin by giving the expressions for the decay amplitudes and
the definitions for the CP asymmetries.  In the
next subsection we collect and discuss the input parameters we use 
to obtain numerical results.  This is followed by a presentation and discussion
of the results.

\subsection{Definitions of Branching Ratios and CP Asymmetries}
\label{BRs_CPAs}

In terms of the coefficients $a_i$ and the factorized matrix elements
defined by
\begin{equation}\label{As}
   A_{M_1\,M_2} = i\frac{G_F}{\sqrt2}\,(m_B^2-m_{M_1}^2)\,
    F_0^{B\to M_1}(m_{M_2}^2)\,f_{M_2},
\end{equation}
the $B\to\pi K$ decay amplitudes are 
\begin{eqnarray}
\label{BKpi}
   {\cal A}(B^-\to\pi^-\bar K^0)
   &=& \lambda_p \left[  a_4^p  
    + \frac{2\,\mu_K}{m_b} a_6^p \right]
    A_{\pi K} \,, \nonumber\\
   - \sqrt2\,{\cal A}(B^-\to\pi^0 K^-)
   &=& \left[ \lambda_u\,a_1 + \lambda_p\,a_4^p  
    + \lambda_p\,\frac{2\,\mu_K}{m_b}\,a_6^p  \right]
    A_{\pi K} + \left[ \lambda_u\,a_2 \right] A_{K\pi}
    \,, \nonumber\\
   - {\cal A}(\bar B^0\to\pi^+ K^-)
   &=& \left[ \lambda_u\,a_1 + \lambda_p\,a_4^p   
    + \lambda_p\,\frac{2\,\mu_K}{m_b}\,a_6^p  \right] A_{\pi K}
    \,, \nonumber\\ 
   \sqrt2\,{\cal A}(\bar{B}^0\to\pi^0 \bar{K}^0)
   &=& {\cal A}(B^-\to\pi^-\bar K^0)
    + \sqrt2\,{\cal A}(B^-\to\pi^0 K^-) \nonumber\\
   &&\mbox{}- {\cal A}(\bar B^0\to\pi^+ K^-) \,.
\end{eqnarray}
In these expressions $a_i\equiv a_i(\pi K)$, $\lambda_p= V_{ps}^* V_{pb}$, and a 
summation over $p \in \{u,c\}$ is implicit in expressions like 
$\lambda_p\,a_i^p$. The last relation is a consequence of isospin symmetry. 

The $B\to\pi\pi$ decay amplitudes are given by
\begin{eqnarray}\label{Bpipi}
   - {\cal A}(\bar B^0\to\pi^+\pi^-)
   &=& \left[ \lambda_u'\,a_1 + \lambda_p'\,a_4^p  
    + \lambda_p'\,\frac{2\,\mu_\pi}{m_b}\, a_6^p \right] A_{\pi\pi}
    \,, \nonumber\\
   - \sqrt2\,{\cal A}(B^-\to\pi^-\pi^0)
   &=& \left[ \lambda_u' (a_1 + a_2) \right] A_{\pi\pi}
    \,, \nonumber\\[0.15cm]
   {\cal A}(\bar{B}^0\to\pi^0\pi^0)
   &=& \sqrt2\,{\cal A}(B^-\to\pi^-\pi^0)
    - {\cal A}(\bar B^0\to\pi^+\pi^-) \,,
\end{eqnarray}
where now $a_i\equiv a_i(\pi\pi)$ and $\lambda_p'=V_{pb} V_{pd}^*$.
The CP conjugate decay amplitudes are obtained from the above by 
replacing $\lambda_p^{(\prime)} \to \left( \lambda_p^{(\prime)}\right)^*$. 
Note that none of these decay modes are dependent on $a_3$ or $a_5$.  These
factors, therefore, play no further role in our discussion.

 CP violation can occur either directly via a difference between CP
conjugate decay rates $(\Gamma(B \to f) \not= \Gamma(\bar{B}\to\bar{f})$ or, for
neutral $B$ mesons, indirectly via $B^0-\bar{B}^0$ mixing.  Accordingly,
we treat the two cases separately.

For the decays $B^- \to \pi^- \bar{K}^0$, $B^- \to \pi^0 K^-$, and
$\bar{B}^0 \to \pi^+ K^-$, the CP asymmetry is time independent and
is defined as
\begin{equation}
\label{ACP1}
{\cal A}_\mathrm{CP}(\bar{f}) = \frac{| {\cal A}(\bar{B} \to \bar{f})|^2 - | {\cal A}(B \to f)|^2 }
				{| {\cal A}(\bar{B} \to \bar{f})|^2 + | {\cal A}(B \to f)|^2}
\end{equation}
where our sign convention is set by defining $\bar{B} = \bar{B}^0,B^-$ as an initial state 
containing a $b$ quark, and $B = B^0,B^+$ as containing an initial $b$ antiquark. 
This CP asymmetry vanishes in the limit of naive factorization, and first occurs
at order $\alpha_s$ in the BBNS formalism.
As such, it can be calculated to order $\alpha_s^2 \beta_0$ with knowledge of only
the next-to-leading order Wilson coefficients.

For the neutral $B$ meson decays to final states $f$ for which
there are interference effects between
$B^0 \to f$ and $B^0 \to \bar{B}^0 \to f$, the resulting CP
asymmetry is time dependent
\begin{equation}
\label{ACP2}
{\cal A}_\mathrm{CP}(t, \bar{f}) = \frac{| {\cal A}(\bar{B}^0(t) \to \bar{f})|^2 - | {\cal A}(B^0(t) \to f)|^2 }{| {\cal A}(\bar{B}^0(t) \to \bar{f})|^2 + | {\cal A}(B^0(t) \to f)|^2}.
\end{equation}
In this paper the modes which fall into this
class are $\bar{B}^0 \to \pi^0 K_S$, $\bar{B}^0 \to \pi^+ \pi^-$,
and $\bar{B}^0 \to \pi^0 \pi^0$.
This asymmetry is often written as
\begin{equation}
\label{time_dep_cpasym}
{\cal A}_\mathrm{CP}(t, \bar{f}) = \mathcal{A}_{\bar{f}} \cos(\Delta m t) -
        \mathcal{S}_{\bar{f}} \sin(\Delta m t)
\end{equation}
where
${\cal A}_{\bar{f}}$ characterizes the direct CP violation due to interference of different
diagrams contributing to the decay,
and $\mathcal{S}_{\bar{f}}$ measures the indirect CP violation which originates from mixing between the
$B^0$ and $\bar{B}^0$ initial states.  Measuring the time dependence of the CP asymmetry
allows one to separate the contributions of these two mechanisms.

Similar to the case of the time independent CP asymmetry above,
$\mathcal{A}_{\bar{f}}$ first occurs at order $\alpha_s$, and
can be calculated with knowledge of only the leading order Wilson coefficients.
Note that $\mathcal{A}_{\bar{f}}$ is simply the time dependent asymmetry
evaluated at $t=0$, and is given by (\ref{ACP1}).
$\mathcal{S}_{\bar{f}}$ on the
other hand is non--zero in naive factorization,
and its determination requires knowledge of the
NNLO Wilson coefficients.
As such, we will not
consider $\mathcal{S}_{\bar{f}}$ any further in this paper.

If we write the Feynman amplitudes (\ref{BKpi})-(\ref{Bpipi}) in
the form ${\cal A} = \lambda_u^{(\prime)} u + \lambda_c^{(\prime)} c$
and decompose the two terms into perturbative contributions
\begin{eqnarray}
u &=& u_0 + \frac{\alpha_s}{4\pi} u_1 +\frac{\alpha_s^2}{(4\pi)^2}\beta_0 u_2 \nonumber \\
c &=& c_0 + \frac{\alpha_s}{4\pi} c_1 +\frac{\alpha_s^2}{(4\pi)^2}\beta_0 c_2
\end{eqnarray}
then we have
\begin{equation}
\label{AcpNonRGI1}
{\cal A}_\mathrm{CP}^\mathrm{Dir} = -2 \mbox{Im}[\lambda_u^{(\prime)}(\lambda_c^{(\prime)})^*]
\frac{\mbox{Im}[u^* c]}{|\lambda_u^{(\prime)}|^2 |u|^2
+ |\lambda_c^{(\prime)}|^2 |c|^2
+ 2 \mbox{Re}[\lambda_u^{(\prime)}(\lambda_c^{(\prime)})^*]
\mbox{Re}[u^* c]
}.
\end{equation}
We
can expand $\mathcal{A}_\mathrm{CP}^\mathrm{Dir}$ to order $\mathcal{O}(\alpha_s^2\beta_0)$
to obtain
\begin{eqnarray}
\label{AcpNonRGI3}
{\cal A}_\mathrm{CP}^\mathrm{Dir} &=&
        \frac{2 \mbox{Im}[ \lambda_u^{(\prime)}(\lambda_c^{(\prime)})^* ]}
        {(\lambda_u^{(\prime)}u_0 + \lambda_c^{(\prime)}c_0)
        ((\lambda_u^{(\prime)})^* u_0 +
        (\lambda_c^{(\prime)})^* c_0)} \nonumber \\
&\,& \hspace{-0.8cm} \times \left\{ \frac{\alpha_s}{4 \pi} \left(u_0 \mbox{Im}[c_1] -c_0 \mbox{Im}[u_1] \right)
        + \frac{\alpha_s^2}{(4\pi)^2}\beta_0 \left(u_0 \mbox{Im}[c_2]-c_0 \mbox{Im}[u_2]\right) \right\}
\end{eqnarray}
where we have used the fact that $u_0$ and $c_0$ are real.  Note that to
this order the direct CP asymmetry is not sensitive to the real part of
the perturbative corrections.  Note also that, as anticipated, we require only the next-to-leading
order behaviour of the Wilson coefficients.

\subsection{Comparison to Previous Work}

As mentioned earlier, the calculations presented in this paper are
similar to calculations presented earlier by Neubert and Pecjak in \cite{Neubert:2002ix}.
Performing a renormalon analysis, they estimated both perturbative and
power corrections in the context of the large-$\beta_0$ limit.
This limit is a way of organizing the perturbative expansion that differs
from what is typically done in renormalization group improved
(RG-improved) perturbation theory.
The most important difference is the power counting.
Rather than expanding in the strong coupling,
$\beta_0$ is taken to be large and one expands
in powers of $1/\beta_0$.
$\alpha_s$ is still considered to be a small parameter in this limit and scales
like $\alpha_s \sim \mathcal{O}(1/\beta_0)$.
In practice the large-$\beta_0$ scaling is implemented
by switching to a rescaled coupling $b(\mu)$ related to the leading order running of
$\alpha_s$:
\begin{equation}
\alpha_s(\mu)\to b(\mu)=\beta_0\frac{\alpha_s(\mu)}{4\pi}=
\frac{1}{\log(\mu^2/\Lambda_{\overline{\mathrm{MS}}}^2)},
\end{equation}
where $b(\mu)\sim\mathcal{O}(1)$.
In contrast to RG-improved perturbation theory, $\log(M/\mu)$ is
of $\mathcal{O}(1)$ in the large-$\beta_0$ limit.
Furthermore, before one expands in $1/\beta_0$, all
occurrences of $n_f$ are replaced by $n_f\to -3\beta_0/2$.

This unusual counting scheme forces one to sum certain classes of diagrams to all orders
in perturbation theory.
The fermion bubbles of Fig. \ref{fig:gluons} are a special case as they scale as
$\alpha_s \beta_0 \sim \mathcal{O}(1)$ after the replacement $n_f\to -3\beta_0/2$.
Thus one must sum an infinite number of fermion bubbles into gluon propagators.
The use of such summations is a common technique in renormalon analyses
\cite{Beneke:1998ui}.

In \cite{Neubert:2002ix}, in order to calculate the $a_i$ to subleading order,
both the hard scattering kernels and the Wilson coefficients
had to be calculated
to NLO in the large-$\beta_0$ limit.
As the hard scattering kernels are $\mathcal{O}(1)$, the Wilson
coefficients had to be calculated to $\mathcal{O}(1/\beta_0)$.
Because they are determined by matching at the weak scale ($\mu=m_W$) and
running down to the scale of the decay ($\mu \sim m_b$), it was necessary
to have the $\mathcal{O}(1/\beta_0)$ matching as well.
However, it was argued by the authors that the difference between
the NLO matching and $\mathcal{O}(1/\beta_0)$ matching was negligible so that
the currently known one-loop (NLO) matching was sufficient.

To run the Wilson coefficients correctly, the elements of the anomalous dimension matrix
had to be determined to the appropriate order.
Because the current-current operators enter at $\mathcal{O}(1)$ in the matching,
pieces of the
anomalous dimension matrix which depend on these
operators were determined to $\mathcal{O}(1/\beta_0)$.
Current-current operators affect the running of both the penguin operators
and the current-current operators themselves.
For the penguin operators, their effect can be determined from the LO anomalous
dimension matrix in \cite{Buchalla:1995vs}.  For the current-current operators
this has been calculated in \cite{Pott:1997eu}.
Penguin operators enter at order $\mathcal{O}(1/\beta_0)$ at the matching scale.
As such, to calculate the elements of the anomalous dimension matrix which depend
on the penguin operators,
it was only necessary to calculate the diagrams to $\mathcal{O}(1)$.
Unlike the current-current operators,
penguin diagrams in the effective theory can be of $\mathcal{O}(1)$ because of
factors of $n_f$ which occur in fermion loops.

In contrast, the calculations we perform in this paper are in the context of the usual
RG-improved perturbation theory, where $\log(M/\mu)\sim\mathcal{O}(1/\alpha_s)$,
and one calculates order by order in $\alpha_s$.
We have calculated the hard scattering kernels to $\mathcal{O}(\alpha_s^2 \beta_0)$.
This corresponds to inserting a single fermion bubble into the gluon propagators of the
$\mathcal{O}(\alpha_s)$ diagrams and replacing $n_f\to -3\beta_0/2$.
Both the previous authors and ourselves had to decide what to do about the factors of $n_f$
which appear in the penguin diagrams. The factors of $n_f$ which enter from the fermion bubble
have corresponding diagrams with gluon and ghost loops which justify their replacement,
but these diagrams are not present for other factors of $n_f$ which emerge from penguin diagrams.
The previous authors took two different approaches to this problem and considered the cases
where they either replaced $n_f\to -3\beta_0/2$
or they left these factors of $n_f$ alone.  They achieved better results from the
second approach, which is physically better justified.
We choose to use only this latter approach in this paper.

To further aid in our calculations, we choose to calculate quantities that first occur
at $\mathcal{O}(\alpha_s)$ in perturbation theory.
It is easily understood that such quantities 
require only the NLO Wilson coefficients from
\cite{Buchalla:1995vs}.
The leading order terms in the Wilson coefficients
sum logs of the form $\alpha_s^n \log^n(M/\mu)$ and scale as $\mathcal{O}(1)$.
The NLO terms sum logs of the form $\alpha_s^{n+1} \log^n(M/\mu)$ and scale as
$\mathcal{O}(\alpha_s)$. Because a two loop calculation is
necessary to calculate these NLO terms they may contain factors of $\beta_0$ in the form
\begin{equation}
\alpha_s^{n+2} \beta_0 \log^{n+1}(M/\mu) \sim \alpha_s \beta_0.
\end{equation}
If we were only after a $\mathcal{O}(\alpha_s)$ result,
we would need the $\mathcal{O}(1)$ (LO) Wilson coefficient and the $\mathcal{O}(\alpha_s)$
hard scattering kernel.  At $\mathcal{O}(\alpha_s^2 \beta_0)$ however, we need
not only the $\mathcal{O}(1)$
Wilson coefficient and the $\mathcal{O}(\alpha_s^2 \beta_0)$ hard scattering kernel, but also
the $\mathcal{O}(\alpha_s)$ (NLO) Wilson coefficient and the $\mathcal{O}(\alpha_s)$
hard scattering kernel.  Since we only need the piece of the NLO Wilson coefficient
proportional to $\beta_0$ we are effectively keeping some unnecessary higher order pieces.
Either way, we require only the NLO Wilson coefficients.

An important effect of the differences between the two approaches is reflected in the different
contributions to the hard scattering kernels which must be calculated.
In order to calculate the coefficients (\ref{ai}), we need many hard scattering and vertex contributions.
Neubert and Pecjak needed fewer of these contributions, but those they did need
were needed to all orders in $\alpha_s^{n+1} \beta_0^n$.
This difference has important phenomenological effects.
The contributions we include are sensitive to unknown non-perturbative parameters.
As we will see, these parameters can introduce a large uncertainty in
various observables.

\subsection{Input Data}

The numerical analysis in section \ref{results} requires various parameters as
theoretical inputs.  In this section we collect these input parameters
together.

\subsubsection{Model Independent Parameters}
\label{ModelIndependentParameters}

For the running coupling $\alpha_s(\mu)$ we use
\begin{equation}
\alpha_s(\mu) = \frac{4\pi}{\beta_0 \ln(\mu^2/\lqcd^2)} \left[ 1 - \frac{\beta_1}{\beta_0^2}
	\frac{\ln( \ln(\mu^2/\lqcd^2))}{\ln(\mu^2/\lqcd^2)} \right]
\end{equation}
where, in terms of the number of colors $N_c$ and flavours $n_f$,
\begin{equation}
\beta_0 = \frac{11 N_c - 2 n_f}{3},\hspace{1cm}\beta_1 = \frac{34 N_c^2}{3}
	- \frac{10 N_c n_f}{3} - 2 C_F n_f,\hspace{1cm}C_F = \frac{N_c^2 -1}{2 N_c}.
\end{equation}
We take $\lqcd = 223\,\mathrm{MeV}$, which is equivalent to running with
$n_f=5$ flavours from $\alpha_s(M_Z) = 0.1185$.

The Wilson coefficients calculated to NLO in QCD are shown in
Table \ref{table:wilsonCoefficients}.
\begin{table}
\caption{Numerical values of the Wilson coefficients $C_{i}$
	in the NDR scheme at NLO,
	in units of $10^{-3}$. We have used the input
	parameters $\lqcd=223\,\mathrm{MeV}$, $m_t = 174\,\mathrm{GeV}$,
	$m_b = 4.2\,\mathrm{GeV}$, and  $m_W = 80.4\,\mathrm{GeV}$.
	The soft scales are defined using $\Lambda_h = 500\,\mathrm{MeV}$. }
 \label{table:wilsonCoefficients}
\begin{center}
 \begin{tabular}{|l|rrrrrr|} \hline
 & \multicolumn{6}{c|}{$\mu$}\\
\cline{2-7}\\[-3ex]
  & $\sqrt{\Lambda_h m_b/2}$ & $\sqrt{\Lambda_h m_b}$ & $\sqrt{\Lambda_h 2 m_b}$ & $m_b/2$ & $m_b$ & $2 \, m_b$ \\ \hline
\input{wilsonCoefficients}
\hline
\end{tabular}
\end{center}
\end{table}

We choose to work in the Wolfenstein parametrization
\begin{equation}
V_\mathrm{CKM}=\left(\begin{array}{ccc}
  1 - \frac{\lambda^2}{2} &  
  \lambda &
  A \lambda^3 (\rho - i \eta) \\
  -\lambda&
  1 - \frac{\lambda^2}{2} &
  A \lambda^2 \\
  A \lambda^3 (1 - \rho - i \eta) &
  -A \lambda^2 &
  1 \end{array}\right) + {\cal O}(\lambda^4).
\end{equation}
In the analysis below we will sometimes plot observables as a 
function of the unitarity angle $\gamma = \mathrm{Arg}[V_{ub}^*]$.
In that case we write $(\rho - i \eta) = \sqrt{\rho^2 + \eta^2} e^{-i \gamma}$.
We take the numerical values of the CKM parameters from a recent
global fit
\cite{Ali:2003te}: $A = 0.83 \pm 0.04$,
$\lambda = 0.2224 \pm 0.0020$, $\sqrt{\rho^2 + \eta^2} = 0.398 \pm 0.040$,
and $\gamma = (64 \pm 11)^o$.

 For the $B$ meson lifetimes we use the PDG values \cite{Eidelman:2004wy}:
$\tau(\bar{B}^0) = 1.536\pm 0.014\,ps$ and $\tau(B^\pm) = 1.671 \pm 0.018\,ps$.
Our quark pole masses are $m_b=4.2\,\mathrm{GeV}$, $m_c=1.3\,\mathrm{GeV}$, and 
we set $m_{s,u,d} = 0$.
Finally, for the `chiral enhancement' factor defined in (\ref{chiral}), which is a renormalization
scale dependent quantity, we use \cite{Du:2001hr}
\begin{equation}
\mu_P(m_b/2) = 0.85 \frac{m_b}{2}, \hspace{1cm} \mu_P(m_b) = 1.14 \frac{m_b}{2}, \hspace{1cm}
	\mu_P(2 m_b) = 1.42 \frac{m_b}{2}
\end{equation}
for both $P=\pi,K$.

\subsubsection{Model Dependent Parameters}
\label{ModelDependentParameters}
 
 The matrix element for a $B$ transition to a pseudoscalar state $M$
is given by
\begin{equation}
\langle M(q) | \bar{q} \gamma^\mu b | \bar{B}(h) \rangle = F_+^{B \to M} (p^2) (h^\mu + q^\mu ) + \left[ F_0^{B \to M}(p^2) 
	-  F_+^{B \to M} (p^2) \right] \frac{ m_B^2 - m_M^2 }{p^2} p^\mu 
\end{equation}
where the momentum transfer is $p = h - q$. In practice this matrix
element is always contracted with one of the meson momenta, and using 
the Dirac equation it is always possible to write these contractions in
terms of $\langle M(q) | \bar{q} \, \xslash{p} \, b | \bar{B}(h) \rangle = 
F_0^{B \to M}(p^2)(m_B^2-m_M^2)$, so that dependence on $F_+$ drops out.
Since we are studying mesons with mass small compared to the 
$B$ mass, we need consider only the point $F_0^{B \to M}(0)$.  
Estimates of this quantity have been made from QCD light--cone sum rules
\cite{Khodjamirian:2000ds, Ball:2001fp, Ball:2004ye}, relativistic quark models
\cite{Melikhov:2000yu}, and lattice calculations \cite{Abada:2000ty}, 
with good agreement between the various methods.
Numerically we take
\begin{equation}
\label{formfactorvalue}
F_0^{B \to K}(0) = \frac{f_K}{f_\pi} F_0^{B \to \pi}(0); \hspace{1cm}   
F_0^{B \to \pi}(0) = 0.258 \pm 0.031\,\mathrm{GeV}.
\end{equation}

The decay constants $f_M$ are defined by
\begin{equation}
\langle M(p) | \bar{q} \gamma^\mu \gamma_5 q | 0 \rangle = -i f_M p^\mu. 
\end{equation}
In our data analysis we will take 
\begin{equation}
\label{decayconstantvalue}
f_\pi = 0.1307\pm0.0004\,\mathrm{GeV} \mbox{\cite{Eidelman:2004wy},\hspace{0.2cm}}
f_K = 0.1598\pm0.0016\,\mathrm{GeV} \mbox{\cite{Eidelman:2004wy}, \hspace{0.2cm}}
f_B = 0.180\pm0.040\;\mathrm{GeV} \mbox{\cite{Abada:1999xd}}.
\end{equation}

The general decomposition of the LCDAs
$\Phi_M$ has been given earlier (\ref{wfdecomposition}) in terms of 
the parameters $\alpha_i^M$, and throughout section \ref{perturbative_corrections} we stated our
results in terms of these parameters. A variety of phenomenological and
sum rule estimates have been made for these parameters \cite{Braun:1988qv,Chernyak:1983ej,rady,Braun:2004vf}, 
and we adopt the values
\begin{eqnarray}
\label{gegmoments}
\alpha_1^\pi &=& 0, \hspace{3.4cm} \alpha_2^\pi = 0.1 \pm 0.3, \nonumber \\
\alpha_1^K &=& 0.10 \pm 0.12, \hspace{2cm} \alpha_2^K = 0.1 \pm 0.3.
\end{eqnarray}
Owing to their nonperturbative origin all of these parameters are rather
poorly known, and this is indicated by the conservative error estimates.
The exception to this rule is $\alpha_1^\pi$ which deviates from 
zero only by $SU(2)$ breaking effects.

Little is known about the LCDA for the light quark in the $B$ meson.
Accordingly, in section \ref{sec:hs_diagrams} 
we, following Ref. \cite{bbns3}, parametrized the integrals
over $\Phi_B$ that we encountered:
\begin{equation}
\label{lambda_defs}
\int_0^1\,d\xi \frac{\Phi_B(\xi)}{\xi} = \frac{m_B}{\lambda_B} \sim \frac{m_B}{\Lambda_h}, \hspace{1cm}
\int_0^1\,d\xi \frac{\ln\xi \; \Phi_B(\xi)}{\xi} = \frac{m_B}{\tilde{\lambda}_B} 
	\sim \frac{m_B}{\Lambda_h} \ln\left( \frac{\Lambda_h}{m_B} \right)
\end{equation}
where we take the soft scale to be $\Lambda_h = 500$ MeV. 
In our numerical analysis we assign a $100\%$ uncertainty to these integrals,
varying $0 < m_B/\lambda_B < 2 (m_B/\Lambda_h)$ and $0 < -m_B/\tilde{\lambda}_B
< 2 (m_B/\Lambda_h \ln(m_B/\Lambda_h))$ with a uniform probability distribution.

 Also in section \ref{sec:hs_diagrams} we saw that integrals over the twist--3 
LCDAs required the introduction of two other parameters
\begin{equation}
\label{X_defs}
\int_0^1\,dy \frac{\Phi_p^{M_1}(y)}{\bar{y}} = X_H^{M_1}, \hspace{1cm}
\int_0^1\,dy \frac{\ln\bar{y}\;\Phi_p^{M_1}(y)}{\bar{y}} = \tilde{X}_H^{M_1}.
\end{equation}
The approximate magnitude of these parameters can be estimated by power
counting, but in general they can be complex.  Therefore, following Ref. \cite{bbns3} 
we write them as
\begin{equation}
\label{Xparametrize}
X_H^{M_1} = \left( 1 + \rho_H e^{i \phi_H} \right) 
	\ln\left(\frac{m_B}{\Lambda_h}\right), \hspace{1cm}  
\tilde{X}_H^{M_1} = \left( 1 + \tilde{\rho}_H e^{i \tilde{\phi}_H} \right)
	\mbox{Li}_2\left( \frac{-m_b}{\Lambda_h} \right).
\end{equation} 
In the numerical analysis below we vary $0 < \rho_H,\tilde{\rho}_H < 2$ and allow
the phases $\phi_H,\tilde{\phi}_H$ to take arbitrary values.  

In the numerical analysis which follows, our central values 
are obtained by setting all of the input parameters at the 
center of their ranges.  For the arbitrary
phases in $X_H$ and $\tilde{X}_H$ we must choose a particular
value for our 'central value'.  
We choose $\phi_H,\tilde{\phi}_H = \pi/2$, which 
makes the real and 
imaginary parts of the central values of $X_H$ and $\tilde{X}_H$ of equal 
magnitude.

\subsection{Results}
\label{results}

The main numerical results of this paper are presented in
Table \ref{table:cpasymmetries} which
shows the results for the direct CP asymmetries ${\cal A}_\mathrm{CP}^\mathrm{Dir}$.
After a general discussion of the results,
we include a more detailed discussion of the
particular asymmetry ${\cal A}_\mathrm{CP}^\mathrm{Dir}(\pi^+ \pi^-)$.
For each quantity we first state
the prediction at ${\cal O}(\alpha_s)$,
then the ${\cal O}(\alpha_s^2 \beta_0)$ correction and, finally, the sum.
We also state the results
at three different renormalization scales.

In addition,
we estimate the uncertainties 
arising from the model dependent parameters discussed in section
\ref{ModelDependentParameters}.  We divide these parameters into
three groups.  The Gegenbauer moments $\alpha_{1,2}$ which 
parametrize the shape of the LCDAs, are varied with the 
$1\,\sigma$ error bars given in (\ref{gegmoments}). The $B$ decay formfactor
$F_0^{B\to\pi}(0)$ and $B$ meson
decay constant $f_B$ are varied with the $1\,\sigma$ error bars
given in (\ref{formfactorvalue}) and (\ref{decayconstantvalue}), respectively.
Finally, the parameters
arising from the spectator scattering graph $(\lambda_B,
\tilde{\lambda}_B, X_H, \tilde{X}_H)$ are varied with equal
probability over the ranges given in section \ref{ModelDependentParameters} above.
These   
three groups are labelled LCDA $(\alpha_{1,2}^{\pi,K})$, FF $(f_B,F_0^{B\to\pi}(0))$,
and SPEC $(\lambda_B,\tilde{\lambda}_B, X_H, \tilde{X}_H)$ in our Table.
The sets of parameters are varied independently, and 
in all cases the uncertainties we give are $1\,\sigma$ 
standard deviations.  
Note that there are additional sources of uncertainty we do not consider,
such as dependence on the CKM
matrix elements and quark masses.  
Our analysis does, however, give
insight into the relative size of perturbative corrections
and nonperturbative uncertainties. 

\subsubsection{CP asymmetries}

Table \ref{table:cpasymmetries} contains our results for the
direct CP asymmetries ${\cal A}_\mathrm{CP}^\mathrm{Dir}$.
It should be noted that the values in Table \ref{table:cpasymmetries} do not take into account
the contribution from weak annihilation diagrams and as such should not be taken as rigorous
predictions of the BBNS method.  They are however valid for their purpose of studying the
perturbative behaviour of the formalism.
Notice that the asymmetry ${\cal A}_\mathrm{CP}(\pi^- \pi^0)$ is not
shown in the table; as is clear from the definitions (\ref{Bpipi}), the
amplitude for $B^- \to \pi^- \pi^0$ has only one weak phase and
therefore the asymmetry
for this mode is zero (up to small electroweak corrections
which we have neglected).
The final column of Table \ref{table:cpasymmetries} gives the current experimental values for the CP asymmetries.
They are from HFAG, Summer  2005 compilation \cite{hfag}, apart from the observables where different experiments do not agree, in which case the errors are inflated according to the PDG prescription \cite{Eidelman:2004wy}.
\begin{table}
\caption{Numerical results for direct CP asymmetries ${\cal A}_\mathrm{CP}^\mathrm{Dir}$,
        expressed in percent.
	We present the ${\cal O}(\alpha_s)$ and
	subleading ${\cal O}(\alpha_s^2 \beta_0)$ perturbative corrections.
	Partial error estimates whose meaning is explained in the text and experimental values are also presented.}
\label{table:cpasymmetries}
\begin{center}
 \begin{tabular}{|c|lrr|r|lll|c|} \hline
 Decay Mode & $\mu$ & ${\cal O}(\alpha_s)$ & ${\cal O}(\alpha_s^2 \beta_0)$ & Total & \multicolumn{3}{c|}{Error Estimates} & Experiment\\
& & & & & LCDA & FF & SPEC & \\ \hline
\input{directcpas}
\hline
\end{tabular}
\end{center}
 \end{table}

Though the
relative sizes of the perturbative contributions at
${\cal O}(\alpha_s)$ and ${\cal O}(\alpha_s^2 \beta_0)$
in Table \ref{table:cpasymmetries} are quite sensitive to the
renormalization scale, it is generally true that the
two contributions are of roughly the same size.
This may be understood as follows:
the asymmetries are dominated by the
contributions from the penguin diagrams, and
$\alpha_s \beta_0 | \tilde{P}/P |/4\pi \sim 1$,
where $P$ and $\tilde{P}$ refer to the
penguin functions
defined in section \ref{sec:penguin_diagrams}.

There is no reduction in the renormalization scale
dependence of the asymmetries after adding the
${\cal O}(\alpha_s^2 \beta_0)$ terms.  This
behaviour follows from our previous remarks:
at a given scale, the ${\cal O}(\alpha_s^2 \beta_0)$
contributions are numerically similar to the
${\cal O}(\alpha_s)$ contributions.  The sum,
therefore, follows the pattern established
at ${\cal O}(\alpha_s)$.

The next to last columns of Table \ref{table:cpasymmetries}
show the sensitivity of the asymmetries to the
three classes of parameters defined above.
The dominant uncertainty for most of the modes is
due to the light cone distribution amplitudes (LCDA).
These parameters are similar in size for each decay mode
(see (\ref{gegmoments})), and the asymmetry is proportional to
them.
Consequently the size of the uncertainty in this
column scales roughly with the size of the asymmetry
itself.

Note also that three of the modes --
$\pi^0 K^-$, $\pi^0 K_S$, and $\pi^0 \pi^0$ --
are particularly sensitive to the form
factor (FF) and spectator scattering (SPEC) parameters.
Unlike other decay modes, the FF and SPEC parameters in these modes
are proportional to the Wilson coefficient $C_1$.
The strong sensitivity in these modes is simply a reflection of the large
size of $C_1$ in comparison to the other Wilson coefficients.

On the other hand, the asymmetry ${\cal A}_\mathrm{CP}^\mathrm{Dir}(\pi^- \overline{K}^0)$
has no dependence on the form factors or spectator scattering parameters.
This may be understood by examining our master formula for the CP asymmetry,
Eq. (\ref{AcpNonRGI3}),
\begin{equation}
\label{AcpUC}
{\cal A}_\mathrm{CP}^\mathrm{Dir} \propto \mbox{Im}[u^*\,c] =
\frac{\alpha_s}{4\pi} \left( u_0 \mbox{Im} [ c_1 ] - c_0 \mbox{Im} [ u_1 ] \right) +
\frac{\alpha_s^2}{(4\pi)^2}\beta_0 \left( u_0 \mbox{Im} [ c_2 ] - c_0 \mbox{Im} [ u_2 ] \right).
\end{equation}
Because of the particular form of the amplitude for $B^- \to \pi^- \overline{K}^0$
shown in (\ref{BKpi}),
$u_0 = c_0$ and Im$[u_i]$ differs from Im$[c_i]$ only by QCD penguin contributions.
The result
is that in (\ref{AcpUC}) only the QCD penguins contribute to the asymmetry. Consequently
this mode is insensitive to most of the model dependence in the BBNS framework.

It is important to note that our values of the CP asymmetries at $\mathcal{O}(\alpha_s)$ for $B^- \to \pi^0 K^-$ and $\bar{B}^0 \to \pi^+ K^-$ exhibit a much greater scale dependence than those of Beneke and Neubert \cite{Beneke:2003zv}.
In our calculation of these asymmetries we keep the real parts of our amplitudes to only LO in $\alpha_s$ and it is the large scale dependence of these real parts that leads to the large scale dependence of our asymmetries.
Beneke and Neubert on the other hand keep the real parts of their amplitudes to NLO and it is these higher order terms that result in their reduced scale dependence.

Because of their different focus, it is difficult to compare our results to those of
Neubert and Pecjak \cite{Neubert:2002ix}.  These authors were primarily interested
in estimating the size of non-perturbative corrections.
Although they did calculate some leading perturbative corrections, 
they sought only to compare the size of these corrections to their estimate of the
power corrections.
As such, almost all of the
parameters we chose to vary, including
the renormalization scale, the LCDAs, as well as the form factors and decay
constants, they simply held fixed, and no estimate of their induced uncertainties
was made.
They did calculate the CP asymmetries for the $\bar{B}^0 \to \pi^+ K^-$ and
$B^- \to \pi^0 K^-$ decay modes.  Their results are
consistent with our own, to within our large uncertainties.
Perhaps the most significant comparison concerns the size of the
subleading corrections.
They found, as we did, that the subleading corrections are substantial
and can be almost as large as the leading order result.

\begin{figure}[htbp]
\centerline{\scalebox{1.0}{\includegraphics{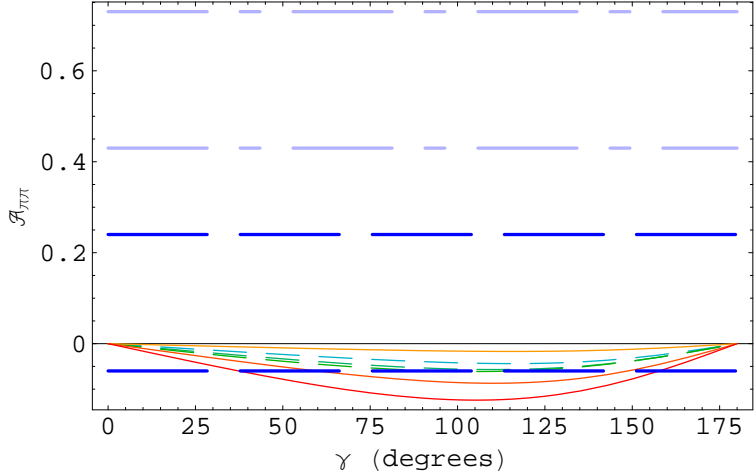}}}
\caption{The CP violating quantity ${\cal A}_{\pi\pi}$ as a function of the unitarity angle
$\gamma$.
The three short--dashed curves are the prediction at order ${\cal O}(\alpha_s)$,
while the three solid curves
include the perturbative corrections up to ${\cal O}(\alpha_s^2 \beta_0)$.
The lines in each set correspond to the three different renormalization scales
$\mu=m_b/2$, $\mu=m_b$ and $\mu=2m_b$.
The heavy dashed and dot-dashed horizontal lines are the $1 \sigma$ experimental
uncertainties for BABAR and BELLE, respectively.}
\label{fig:BppA}
\end{figure}

Recently new measurements of the CP asymmetry in $\bar{B}^0 \to \pi^+ \pi^-$
were released by the  BABAR \cite{Aubert:2005av} and BELLE \cite{Abe:2005dz}
collaborations.
In Figure \ref{fig:BppA} we show our results for the
direct CP violation parameter ${\cal A}_{\pi\pi}$ defined in
(\ref{time_dep_cpasym})
as a function of the unitarity angle $\gamma$.
The current experimental data for this quantity is
\begin{eqnarray}
\mbox{BABAR} \mbox{\cite{Aubert:2005av}} &:&  \begin{array}{c}
{\cal A}_{\pi\pi} = 0.09 \pm 0.15\mbox{(stat)}\pm 0.04\mbox{(syst)}
\end{array}  \nonumber \\
\mbox{BELLE} \mbox{\cite{Abe:2005dz}} &:&  \begin{array}{c}
{\cal A}_{\pi\pi} = 0.56 \pm 0.12\mbox{(stat)}\pm 0.06\mbox{(syst)},
\end{array}
\end{eqnarray}
The $1 \sigma$ ranges for these measurements are superimposed in
Figure \ref{fig:BppA}.

Figure \ref{fig:BppA} shows that the calculated CP asymmetry has a large
renormalization scale dependence at ${\cal O}(\alpha_s^2 \beta_0)$,
which dominates the uncertainty in the prediction.
Within the large error bars, the experimental results of BABAR are
in fair agreement with these calculations, while the results of BELLE show a several
$\sigma$ deviation.
Clearly, more
work is required on both the experimental and theoretical
sides before any definitive statement can be made about
the success of the BBNS framework in this context.
For instance, the contributions from power--suppressed
annihilation diagrams should be included, as they
are known to have a large effect on strong phases
\cite{bbns3,Keum:2000ph}.

\section{Conclusions}

In this paper we have calculated perturbative corrections to
$B \to \pi \pi, \pi K$ decays  up to ${\cal O}(\alpha_s^2 \beta_0)$ in the QCD
factorization formalism, including `chirally enhanced' power corrections but
neglecting other corrections entering formally at ${\cal O}(\lqcd/m_b)$.
We have included contributions from non--factorizable vertex diagrams,
QCD penguin diagrams, and spectator scattering diagrams.  In all cases
we have derived analytic expressions for the hard scattering kernels
for general light--cone quark momentum distribution functions.

We have used these analytic results to study the direct CP
asymmetries for a set of phenomenologically interesting
decay modes.
We focused on the behaviour of perturbation theory for
this observable, and we
estimated the uncertainties due to model
dependent parameters.

For the direct CP asymmetries ${\cal A}_\mathrm{CP}^\mathrm{Dir}$,
we found that contributions at
${\cal O}(\alpha_s^2 \beta_0)$ are comparable to those
at ${\cal O}(\alpha_s)$ for all the modes.  This conclusion is in
agreement with the results of \cite{Neubert:2002ix}, which indicated a
large perturbative correction between one-loop and two-loop order
in the large-$\beta_0$ limit.
As well, we found a
very strong dependence on the renormalization scale; in some
cases the asymmetry varies over an order of magnitude.

For all modes, with the exception of ${\cal A}_\mathrm{CP}^\mathrm{Dir}( \pi^0 \pi^0)$,
the primary uncertainty at a given scale arises from
uncertainty over the shape of the light cone momentum
distribution amplitude.  The uncertainties arising
from form factors and spectator scattering model parameters
are mode dependent and relatively small.
The asymmetry
${\cal A}_\mathrm{CP}^\mathrm{Dir}( \pi^- \overline{K}^0)$ is particularly clean
in the QCD factorization framework, having no dependence on
form factors or spectator scattering parameters.
Finally, we have examined the direct CP violation
parameter $\mathcal{A}_{\pi\pi}$ in the
$\bar{B}^0 \to \pi^+ \pi^-$ decay mode and
have found a large perturbative
correction at ${\cal O}(\alpha_s^2 \beta_0)$.
The result agrees with the current experimental measurements,
though the errors for both theory and experiment are large.

\section{Acknowledgments}
We would like to thank Michael Luke for discussions related to
this project.  This work is supported by the Natural Sciences and Engineering 
Research Council of Canada and by the Walter B. Sumner Foundation.

\end{document}

%% file: wilsonCoefficients.tex
$C_1$ & $1258.0$ & $1195.2$ & $1150.4$ & $1147.7$ & $1087.8$ & $1048.6$ \\
$C_2$ & $-474.8$ & $-378.7$ & $-305.3$ & $-300.7$ & $-193.3$ & $-114.4$ \\
$C_3$ & $35.7$ & $27.4$ & $21.6$ & $21.2$ & $13.8$ & $9.0$ \\
$C_4$ & $-77.7$ & $-62.8$ & $-52.0$ & $-51.3$ & $-36.0$ & $-25.0$ \\
$C_5$ & $11.9$ & $12.4$ & $11.9$ & $11.9$ & $9.9$ & $7.7$ \\
$C_6$ & $-118.6$ & $-88.2$ & $-68.5$ & $-67.4$ & $-43.3$ & $-28.3$ \\
$C_{8g}^{\mathrm{eff}}$ & - & - & - & $-169.0$ & $-151.0$ & $-136.0$ \\

%% file: directcpas.tex
 & $m_b/2$ & $0.9$ & $0.1$ & $0.9$ & $\pm{0.1}$ & $\pm{0.0}$ & $\pm{0.0}$ &  \\
${\cal A}_{CP}(\pi^- \overline{K}^0)$ & $m_b$ & $1.0$ & $0.3$ & $1.2$ & $\pm{0.1}$ & $\pm{0.0}$ & $\pm{0.0}$ & $-2.\pm5.$ \\
 & $2m_b$ & $1.2$ & $0.4$ & $1.6$ & $\pm{0.1}$ & $\pm{0.0}$ & $\pm{0.0}$ &  \\\hline
 & $m_b/2$ & $9.1$ & $-1.4$ & $7.7$ & $\pm{2.5}$ & $\pm{0.5}$ & $\pm{1.4}$ &  \\
${\cal A}_{CP}(\pi^0 K^-)$ & $m_b$ & $15.9$ & $8.3$ & $24.2$ & $\pm{3.0}$ & $\,^{+0.6}_{-0.7}$ & $\pm{1.8}$ & $4.\pm4.$ \\
 & $2m_b$ & $29.4$ & $28.9$ & $58.2$ & $\pm{4.1}$ & $\,^{+0.8}_{-0.9}$ & $\pm{2.5}$ &  \\\hline
 & $m_b/2$ & $3.9$ & $-3.1$ & $0.7$ & $\pm{2.2}$ & $\pm{0.0}$ & $\pm{0.1}$ &  \\
${\cal A}_{CP}(\pi^+ K^-)$ & $m_b$ & $9.3$ & $4.8$ & $14.0$ & $\pm{2.3}$ & $\pm{0.0}$ & $\pm{0.0}$ & $-11.5\pm1.8$ \\
 & $2m_b$ & $19.2$ & $20.1$ & $39.3$ & $\pm{3.1}$ & $\pm{0.0}$ & $\pm{0.1}$ &  \\\hline
 & $m_b/2$ & $-3.0$ & $-1.2$ & $-4.2$ & $\pm{0.9}$ & $\pm{0.4}$ & $\pm{1.1}$ &  \\
${\cal A}_{CP}(\pi^0 K_S)$ & $m_b$ & $-3.2$ & $-2.0$ & $-5.2$ & $\pm{1.0}$ & $\pm{0.4}$ & $\pm{1.2}$ & $2.\pm13.$ \\
 & $2m_b$ & $-4.5$ & $-4.3$ & $-8.8$ & $\pm{1.1}$ & $\,^{+0.5}_{-0.4}$ & $\pm{1.4}$ &  \\\hline
 & $m_b/2$ & $-2.6$ & $1.6$ & $-1.0$ & $\pm{1.1}$ & $\pm{0.0}$ & $\pm{0.1}$ &  \\
${\cal A}_{CP}(\pi^+ \pi^-)$ & $m_b$ & $-3.6$ & $-1.9$ & $-5.6$ & $\pm{0.7}$ & $\pm{0.0}$ & $\pm{0.0}$ & $37.\pm23.$ \\
 & $2m_b$ & $-4.2$ & $-4.3$ & $-8.5$ & $\pm{0.5}$ & $\pm{0.0}$ & $\pm{0.0}$ &  \\\hline
 & $m_b/2$ & $68.6$ & $28.8$ & $97.4$ & $\pm{16.0}$ & $\,^{+10.7}_{-12.3}$ & $\pm{33.5}$ &  \\
${\cal A}_{CP}(\pi^0 \pi^0)$ & $m_b$ & $40.7$ & $26.5$ & $67.2$ & $\pm{10.4}$ & $\,^{+6.1}_{-6.9}$ & $\pm{19.2}$ & $28.\pm40.$ \\
 & $2m_b$ & $25.3$ & $25.4$ & $50.7$ & $\pm{5.7}$ & $\,^{+3.0}_{-3.4}$ & $\pm{9.5}$ &  \\\hline

%% file: btoll.bbl
\begin{thebibliography}{99}
%
\bibitem{Cronin-Hennessy:kg}
D.~Cronin-Hennessy {\it et al.}  [CLEO Collaboration],
Phys.\ Rev.\ Lett.\  {\bf 85}, 515 (2000).
%
\bibitem{Asner:2001eh}
D.~M.~Asner {\it et al.}  [CLEO Collaboration],
Phys.\ Rev.\ D {\bf 65}, 031103 (2002)
[arXiv:hep-ex/0103040].
%
\bibitem{Chen:2000hv}
S.~Chen {\it et al.}  [CLEO Collaboration],
Phys.\ Rev.\ Lett.\  {\bf 85}, 525 (2000)
[arXiv:hep-ex/0001009].
%
\bibitem{Aubert:2001hs}
B.~Aubert {\it et al.}  [BABAR Collaboration],
Phys.\ Rev.\ Lett.\  {\bf 87}, 151802 (2001)
[arXiv:hep-ex/0105061].
%
\bibitem{Aubert:2001hk}
B.~Aubert {\it et al.}  [BABAR Collaboration],
Phys.\ Rev.\ D {\bf 65}, 051502 (2002)
[arXiv:hep-ex/0110062].
%
\bibitem{Aubert:2001ap}
B.~Aubert {\it et al.}  [BABAR Collaboration],
arXiv:hep-ex/0109007.
%
\bibitem{Aubert:2002ng}
B.~Aubert {\it et al.}  [BABAR Collaboration],
arXiv:hep-ex/0206053.
%
\bibitem{Aubert:2002jb}
B.~Aubert {\it et al.}  [BABAR Collaboration],
Phys.\ Rev.\ Lett.\  {\bf 89}, 281802 (2002)
[arXiv:hep-ex/0207055].
%
\bibitem{Aubert:2002jj}
B.~Aubert {\it et al.}  [BABAR Collaboration],
arXiv:hep-ex/0207063.
%
\bibitem{Aubert:2002jm}
B.~Aubert {\it et al.}  [BABAR Collaboration],
arXiv:hep-ex/0207065.

\bibitem{Aubert:2004xc}
  B.~Aubert {\it et al.}  [BABAR Collaboration],
  Phys.\ Rev.\ Lett.\  {\bf 93}, 231804 (2004)
  [arXiv:hep-ex/0408017].

\bibitem{Aubert:2005av}
  B.~Aubert {\it et al.}  [BaBar Collaboration],
  arXiv:hep-ex/0501071.

\bibitem{Abe:2001nq}
K.~Abe {\it et al.}  [BELLE Collaboration],
Phys.\ Rev.\ Lett.\  {\bf 87}, 101801 (2001)
[arXiv:hep-ex/0104030].

\bibitem{Abe:2001hs}
K.~Abe {\it et al.}  [BELLE Collaboration],
Phys.\ Rev.\ D {\bf 64}, 071101 (2001)
[arXiv:hep-ex/0106095].

\bibitem{Abe:2004us}
  K.~Abe {\it et al.}  [Belle Collaboration],
  Phys.\ Rev.\ Lett.\  {\bf 93}, 021601 (2004)
  [arXiv:hep-ex/0401029].

\bibitem{Casey:2002yd}
B.~C.~K.~Casey {\it et al.}  [Belle Collaboration],
Phys.\ Rev.\ D {\bf 66}, 092002 (2002)
[arXiv:hep-ex/0207090].

\bibitem{Abe:2005dz}
  K.~Abe {\it et al.}  [Belle Collaboration],
  arXiv:hep-ex/0502035.

\bibitem{Bauer:1986bm}
M.~Bauer, B.~Stech and M.~Wirbel,
Z.\ Phys.\ C {\bf 34}, 103 (1987).
%
\bibitem{Deandrea:1993ma}
A.~Deandrea, N.~Di Bartolomeo, R.~Gatto and G.~Nardulli,
Phys.\ Lett.\ B {\bf 318}, 549 (1993)
[arXiv:hep-ph/9308210].
%
\bibitem{NRSX}
M. Neubert, V. Rieckert, B. Stech and Q.P. Xu, in: {\em Heavy Flavours},
ed.\ A.J. Buras and M. Lindner (World Scientific, Singapore, 1992),
pp.~286.
%
\bibitem{Neubert:1997uc}
M.~Neubert and B.~Stech,
Adv.\ Ser.\ Direct.\ High Energy Phys.\  {\bf 15}, 294 (1998)
[arXiv:hep-ph/9705292].
%
\bibitem{bbns1}
M.~Beneke, G.~Buchalla, M.~Neubert and C.~T.~Sachrajda,
Phys.\ Rev.\ Lett.\  {\bf 83}, 1914 (1999)
[arXiv:hep-ph/9905312].
%
\bibitem{bbns2}
M.~Beneke, G.~Buchalla, M.~Neubert and C.~T.~Sachrajda,
Nucl.\ Phys.\ B {\bf 591}, 313 (2000)
[arXiv:hep-ph/0006124].
%
\bibitem{bbns3}
M.~Beneke, G.~Buchalla, M.~Neubert and C.~T.~Sachrajda,
Nucl.\ Phys.\ B {\bf 606}, 245 (2001)
[arXiv:hep-ph/0104110].
%
\bibitem{Bauer:2001yt}
C.~W.~Bauer, D.~Pirjol and I.~W.~Stewart,
Phys.\ Rev.\ D {\bf 65}, 054022 (2002)
[arXiv:hep-ph/0109045].
%
\bibitem{Bauer:2001cu}
C.~W.~Bauer, D.~Pirjol and I.~W.~Stewart,
Phys.\ Rev.\ Lett.\  {\bf 87}, 201806 (2001)
[arXiv:hep-ph/0107002].
%

\bibitem{Bauer:2002aj}
C.~W.~Bauer, D.~Pirjol and I.~W.~Stewart,
Phys.\ Rev.\ D {\bf 67}, 071502 (2003)
[arXiv:hep-ph/0211069].

\bibitem{Chay:2002vy}
J.~Chay and C.~Kim,
Phys.\ Rev.\ D {\bf 65}, 114016 (2002)
[arXiv:hep-ph/0201197].

\bibitem{Chay:2002mw}
J.~g.~Chay and C.~Kim,
arXiv:hep-ph/0205117.

\bibitem{Beneke:2002ph}
M.~Beneke, A.~P.~Chapovsky, M.~Diehl and T.~Feldmann,
Nucl.\ Phys.\ B {\bf 643}, 431 (2002)
[arXiv:hep-ph/0206152].

\bibitem{Rothstein:2003wh}
  I.~Z.~Rothstein,
  Phys.\ Rev.\ D {\bf 70} (2004) 054024
  [arXiv:hep-ph/0301240].

\bibitem{Chay:2003ju}
  J.~Chay and C.~Kim,
  Nucl.\ Phys.\ B {\bf 680}, 302 (2004)
  [arXiv:hep-ph/0301262].

\bibitem{Bauer:2004tj}
  C.~W.~Bauer, D.~Pirjol, I.~Z.~Rothstein and I.~W.~Stewart,
  Phys.\ Rev.\ D {\bf 70}, 054015 (2004)
  [arXiv:hep-ph/0401188].

\bibitem{Muta:2000ti}
T.~Muta, A.~Sugamoto, M.~Z.~Yang and Y.~D.~Yang,
Phys.\ Rev.\ D {\bf 62}, 094020 (2000)
[arXiv:hep-ph/0006022].
%
\bibitem{Du:2000ff}
D.~s.~Du, D.~s.~Yang and G.~h.~Zhu,
Phys.\ Lett.\ B {\bf 488}, 46 (2000)
[arXiv:hep-ph/0005006].

\bibitem{Du:2001hr}
  D.~s.~Du, H.~u.~Gong, J.~f.~Sun, D.~s.~Yang and G.~h.~Zhu,
  Phys.\ Rev.\ D {\bf 65}, 074001 (2002)
  [arXiv:hep-ph/0108141].

\bibitem{Rhadrons}
K.~G.~Chetyrkin, A.~L.~Kataev and F.~V.~Tkachov,
Phys.\ Lett.\ B {\bf 85}, 277 (1979);

M.~Dine and J.~R.~Sapirstein,
Phys.\ Rev.\ Lett.\  {\bf 43}, 668 (1979);

W.~Celmaster and R.~J.~Gonsalves,
Phys.\ Rev.\ Lett.\  {\bf 44}, 560 (1980).

\bibitem{taudecay}
S.~Narison and A.~Pich,
Phys.\ Lett.\ B {\bf 211}, 183 (1988);

E.~Braaten,
Phys.\ Rev.\ D {\bf 39}, 1458 (1989).
%
\bibitem{btoupert}
M.~E.~Luke, M.~J.~Savage and M.~B.~Wise,
Phys.\ Lett.\ B {\bf 343}, 329 (1995)
[arXiv:hep-ph/9409287];

T.~van Ritbergen,
Phys.\ Lett.\ B {\bf 454}, 353 (1999)
[arXiv:hep-ph/9903226];

M.~Steinhauser and T.~Seidensticker,
Phys.\ Lett.\ B {\bf 467}, 271 (1999)
[arXiv:hep-ph/9909436].
%
\bibitem{Neubert:2002ix}
M.~Neubert and B.~D.~Pecjak,
JHEP {\bf 0202}, 028 (2002)
[arXiv:hep-ph/0202128].
%
\bibitem{Burrell:2001pf}
C.~N.~Burrell and A.~R.~Williamson,
Phys.\ Rev.\ D {\bf 64}, 034009 (2001)
[arXiv:hep-ph/0101190].
%
\bibitem{Becher:2001hu}
T.~Becher, M.~Neubert and B.~D.~Pecjak,
Nucl.\ Phys.\ B {\bf 619}, 538 (2001)
[arXiv:hep-ph/0102219].
%
\bibitem{Buchalla:1995vs}
G.~Buchalla, A.~J.~Buras and M.~E.~Lautenbacher,
Rev.\ Mod.\ Phys.\  {\bf 68}, 1125 (1996)
[arXiv:hep-ph/9512380].

\bibitem{Manohar:1994kq}
A.~V.~Manohar and M.~B.~Wise,
Phys.\ Lett.\ B {\bf 344}, 407 (1995)
[arXiv:hep-ph/9406392].

%
\bibitem{Braun:1988qv}
V.~M.~Braun and I.~E.~Filyanov,
Z.\ Phys.\ C {\bf 44}, 157 (1989)
[Sov.\ J.\ Nucl.\ Phys.\  {\bf 50}, 511.1989\ YAFIA,50,818 (1989)].
%
\bibitem{Braun:1989iv}
V.~M.~Braun and I.~E.~Filyanov,
Z.\ Phys.\ C {\bf 48}, 239 (1990)
[Sov.\ J.\ Nucl.\ Phys.\  {\bf 52}, 126 (1990)].

\bibitem{Beneke:2002bs}
  M.~Beneke,
  Nucl.\ Phys.\ Proc.\ Suppl.\  {\bf 111}, 62 (2002)
  [arXiv:hep-ph/0202056].

\bibitem{Lepage:1980fj}
G.~P.~Lepage and S.~J.~Brodsky,
Phys.\ Rev.\ D {\bf 22}, 2157 (1980).
%
\bibitem{Chernyak:1983ej}
V.~L.~Chernyak and A.~R.~Zhitnitsky,
Phys.\ Rept.\  {\bf 112}, 173 (1984).
%
\bibitem{rady}
S.V.\ Mikhailov and A.V.\ Radyushkin, JETP Lett.\ {\bf 43} 712 (1986);
Sov.\ J.\ Nucl.\ Phys.\ {\bf 49} 494 (1989); Phys.\ Rev.\ D {\bf 45}
(1992) 1754.
%
\bibitem{Buras:1989xd}
A.~J.~Buras and P.~H.~Weisz,
Nucl.\ Phys.\ B {\bf 333}, 66 (1990).

\bibitem{Beneke:1998ui}
M.~Beneke,
Phys.\ Rept.\  {\bf 317}, 1 (1999)
[arXiv:hep-ph/9807443].
%
\bibitem{Pott:1997eu}
N.~Pott,
arXiv:hep-ph/9710503.
%
\bibitem{Ali:2003te}
A.~Ali,
arXiv:hep-ph/0312303.

\bibitem{Eidelman:2004wy}
 S.~Eidelman {\it et al.}  [Particle Data Group],
  Phys.\ Lett.\ B {\bf 592}, 1 (2004).

\bibitem{Khodjamirian:2000ds}
A.~Khodjamirian, R.~Ruckl, S.~Weinzierl, C.~W.~Winhart and O.~I.~Yakovlev,
Phys.\ Rev.\ D {\bf 62}, 114002 (2000)
[arXiv:hep-ph/0001297].

\bibitem{Ball:2001fp}
  P.~Ball and R.~Zwicky,
  JHEP {\bf 0110}, 019 (2001)
  [arXiv:hep-ph/0110115].

\bibitem{Ball:2004ye}
  P.~Ball and R.~Zwicky,
  Phys.\ Rev.\ D {\bf 71}, 014015 (2005)
  [arXiv:hep-ph/0406232].

\bibitem{Melikhov:2000yu}
D.~Melikhov and B.~Stech,
Phys.\ Rev.\ D {\bf 62}, 014006 (2000)
[arXiv:hep-ph/0001113].
%
\bibitem{Abada:2000ty}
A.~Abada, D.~Becirevic, P.~Boucaud, J.~P.~Leroy, V.~Lubicz and F.~Mescia,
Nucl.\ Phys.\ B {\bf 619}, 565 (2001)
[arXiv:hep-lat/0011065].

\bibitem{Abada:1999xd}
A.~Abada, D.~Becirevic, P.~Boucaud, J.~P.~Leroy, V.~Lubicz, G.~Martinelli and F.~Mescia,
Nucl.\ Phys.\ Proc.\ Suppl.\  {\bf 83}, 268 (2000)
[arXiv:hep-lat/9910021].

\bibitem{Braun:2004vf}
  V.~M.~Braun and A.~Lenz,
  Phys.\ Rev.\ D {\bf 70}, 074020 (2004)
  [arXiv:hep-ph/0407282].

\bibitem{hfag}
Heavy Flavor Averaging Group (HFAG) data averages can be found at www.slac.stanford.edu/xorg/hfag/

\bibitem{Beneke:2003zv}
  M.~Beneke and M.~Neubert,
  Nucl.\ Phys.\ B {\bf 675}, 333 (2003)
  [arXiv:hep-ph/0308039].

\bibitem{Keum:2000ph}
Y.~Y.~Keum, H.~n.~Li and A.~I.~Sanda,
Phys.\ Lett.\ B {\bf 504}, 6 (2001)
[arXiv:hep-ph/0004004].

\end{thebibliography}
